\documentclass[pra,twocolumn,reprint]{revtex4-1}

\usepackage[utf8]{inputenc} 
\usepackage{lipsum}
\usepackage{color}
\usepackage{bm}
\usepackage{amssymb,amsmath}
\usepackage{bbold}
\usepackage{hyperref}
\hypersetup{
  colorlinks   = true, %Colours links instead of ugly boxes
  urlcolor     = blue, %Colour for external hyperlinks
  linkcolor    = blue, %Colour of internal links
  citecolor   = blue %Colour of citations
}
\usepackage{makecell}
\usepackage[normalem]{ulem}
\usepackage{graphicx}
\usepackage{pdfpages}
\makeatletter
\AtBeginDocument{\let\LS@rot\@undefined}
\makeatother

\newcommand{ \parO }[1]{\left( #1 \right)}
\newcommand{ \parS }[1]{\left[ #1 \right]}
\newcommand{ \parC }[1]{\left\{ #1 \right\}}
\newcommand{ \dB }[0]{\textbf{b}}
\newcommand{ \Nm }[0]{\mathcal{N}}
\newcommand{ \im }[0]{\mathrm{Im}}
\newcommand{ \ret }[0]{\mathrm{ret}}
\newcommand{ \tr }[0]{\mathrm{Tr}}
\newcommand{ \del }[0]{\partial}
\newcommand{ \id }[0]{\mathbb{1}}
\newcommand{ \upa }[0]{\uparrow}
\newcommand{ \da }[0]{\downarrow}
\newcommand{ \Bv }[0]{\mathbf{B}}
\newcommand{ \xv }[0]{\mathbf{x}}

\newcommand{ \zv }[0]{\mathbf{z}}
\newcommand{ \sigv }[0]{\boldsymbol{\sigma}}

\newcommand{ \tz }[0]{\tau_z}
\newcommand{ \tx }[0]{\tau_x}

\newcommand{ \ra }[0]{\rightarrow}
\newcommand{ \sy }[0]{\sigma_y}

\begin{document}

\title{Majorana bound state engineering via efficient real-space parameter optimization}

\author{Samuel Boutin}
\email{Samuel.Boutin@USherbrooke.ca}
\affiliation{Institut quantique et D\'{e}partement de Physique, Universit\'{e} de Sherbrooke, Sherbrooke, Qu\'{e}bec J1K 2R1, Canada}
\author{Julien Camirand Lemyre}
\affiliation{Institut quantique et D\'{e}partement de Physique, Universit\'{e} de Sherbrooke, Sherbrooke, Qu\'{e}bec J1K 2R1, Canada}
\author{Ion Garate}
\affiliation{Institut quantique et D\'{e}partement de Physique, Universit\'{e} de Sherbrooke, Sherbrooke, Qu\'{e}bec J1K 2R1, Canada}

\date{\today}                                       
%%%%%%%%%%%%%%%%%%%%%%%%%%%%%%%%%%%%%%%%%%%%%%%%%%%%%%%%%%%%%%%%%%%%%%%%%%%%%%%
\begin{abstract}
Recent progress toward the fabrication of Majorana-based qubits has sparked the need for systematic approaches to optimize experimentally relevant parameters for the realization of robust Majorana bound states. 
Here, we introduce an efficient numerical method for the real-space optimization of tunable parameters, such as electrostatic potential profiles and magnetic field textures, in Majorana wires.
Combining ideas from quantum control and quantum transport, our algorithm,
applicable to any noninteracting tight-binding model,
operates on a largely unexplored parameter space and opens new routes for Majorana bound states with enhanced robustness. 
Contrary to common belief, we find that spatial inhomogeneities of parameters can be a resource for the engineering of Majorana bound states. 
\end{abstract}

\maketitle

%%%%%%%%%%%%%%%%%%%%%%%%%%%%%%%%%%%%%%%%%%%%%%%%%%%%%%%%%%%%%%%%%%%%%%%%%%%%%%%
% Introduction
%%%%%%%%%%%%%%%%%%%%%%%%%%%%%%%%%%%%%%%%%%%%%%%%%%%%%%%%%%%%%%%%%%%%%%%%%%%%%%%
% \emph{Introduction.---}
\section{Introduction}\label{sec:intro}
Majorana bound states (MBS) are spatially localized zero-energy modes that exhibit non-abelian exchange statistics.
The recent discovery and characterization of MBS in solid-state devices has established their potential for future fault-tolerant quantum computers~\cite{[{For recent experiments, see e.g. }]Mourik:2012ee,*Deng:2016wt,*Chene2017,*Zhang2018a}.
At present, the leading platform for the study of MBS is a strongly spin-orbit coupled semiconducting nanowire, proximity-coupled to a $s$-wave superconductor and placed in a uniform magnetic field~\cite{Kitaev:2001sp, Oreg:2010xy,[{For reviews, see e.g. }]Pientka2015review,*alicea:2012fj,*Beenakker:2013ve,*AguadoReview,Lutchyn2017}.     
Alternative proposals replacing spin-orbit interactions with spiral magnetic textures generated by adatoms~\cite{Choy2011,Nadj-Perge:2013cr} or arrays of micromagnets~\cite{Kjaergaard:2012tg} are also promising and have been partially realized in experiments~\cite{[{See e.g. }]Nadj-Perge:2014qv,*Ruby2015,*Pawlak2016,*Ruby2017}.

In spite of the aforementioned advances, the current state of knowledge for the realization of MBS is restricted to a small region of parameter space, comprised mainly of translationally-invariant wires.
Systems with nonuniform parameters, 
such as superconducting gaps, magnetic fields and electrostatic potential profiles,
are not analytically tractable beyond a few limiting periodic cases~\cite{Klinovaja:2012, Pientka:2013, Rainis:2014le, Marra2017, DeGottardi:2013a, DeGottardi:2013b, Perez:2017, Hoffman:2016}, and the existing numerical studies~\cite{Tezuka:2012, Poyhonen:2014, Zhang:2016, Moore:2016, Liu:2018} have not been exhaustive. 
Thus, it would be desirable to chart the vast space of tunable experimental parameters beyond the known subregions, not only to find out if inhomogeneities could be a resource for MBS experiments, 
but also to provide new insights for improving Majorana-based qubits~\cite{Plugge:2017dg,Karzig2017,Knapp:2018}.

In this work, we introduce an optimization algorithm that undertakes an efficient search in parameter space for maximally robust MBS which are compatible with experimental constraints. 
The central finding of our work is that the engineering of spatial inhomogeneities increases the parameter space region for robust MBS and significantly enhances the degeneracy of Majorana zero-modes.

Our optimization approach is inspired by the Gradient Ascent Pulse Engineering (GRAPE) algorithm~\cite{Khaneja:2005tw} of quantum optimal control~\cite{[{For reviews, see e.g. }]Glaser:2015ug,*Koch:2016ly,*RabitzReview}, which aims to find the best pulse shapes in time domain to perform a task, such as implementing a logical gate or reaching a desired ground state~\cite{[{See e.g. }]Motzoi:2009ly,*Rahmani:2017fs,*jager2014}. 
We draw an analogy between GRAPE and the Recursive Green's function (RGF) method of quantum transport~\cite{Thouless:1981wq}, which allows to transfer the insights of the former from time domain to real-space domain. This analogy turns out to be key to implement the efficient optimization of parameters for the creation of robust MBS in inhomogeneous quantum wires.

%%%%%%%%%%%%%%%%%%%%%%%%%%%%%%%%%%%%%%%%%%%%%%%%%%%%%%%%%%%%%%%%%%%%%%%%%%%%%%%
% Algorithm and optimal control analogy
%%%%%%%%%%%%%%%%%%%%%%%%%%%%%%%%%%%%%%%%%%%%%%%%%%%%%%%%%%%%%%%%%%%%%%%%%%%%%%%
\section{Real-space analog of optimal control}\label{sec:realspaceOCT}
In quantum optimal control theory, one considers a system with a Hamiltonian $H = H_0 + \sum_k f_k(t) {\cal H}_k$, where $f_k(t)$ are some experimentally controllable time-dependent parameters and $k=1,\dots, p$ labels distinct control fields. 
The control problem can be stated as the maximization of a functional $\Phi[\{f_k(t)\}]$, known as a performance index, which defines the success in accomplishing a desired task.
To make this optimization problem tractable, Khaneja \textit{et al.}~\cite{Khaneja:2005tw} discretized the control functions into piecewise-constant segments $f_k(t_j)$.
The gradient of $\Phi$ with respect to $\parC{f_k(t_j)}$ can then be efficiently calculated by keeping in memory intermediate results of forward-in-time and backward-in-time propagator products computed iteratively (see Appendix~\ref{sec:grape} for a more detailed introduction to GRAPE).
This insight at the core of GRAPE leads to a polynomial speedup (in the number of time steps) of numerical calculations compared to a finite-difference gradient calculation

%%%%%%%%%%%%%%%%%%%%%%%%%%%%%%%%%%%%%%%%%%%%%%%%%%%%%%%%%%%%%%%%%%%%%%%%%%%%%%%
% Model and RGF
%%%%%%%%%%%%%%%%%%%%%%%%%%%%%%%%%%%%%%%%%%%%%%%%%%%%%%%%%%%%%%%%%%%%%%%%%%%%%%%
The piecewise constant approximation of time-domain functions in GRAPE is reminiscent of tight-binding models in condensed matter physics, where space is discretized into a lattice. 
Hence, it is natural to ask whether a real-space analog of GRAPE could be developed to optimize profiles of tunable static experimental parameters in one dimensional wires.
In the following, we pursue this analogy for a wire with Hamiltonian
\begin{equation}
	H = \sum_j \left(\psi_j^\dag h_j   \psi_j
	+ \psi_{j+1}^\dag  u_j  \psi_{j} 
	+ \psi_{j}^\dag u_j^\dag    \psi_{j+1}\right),
	\label{eq:hTB}
\end{equation}
where $j$ is the site index. 
Each site contains $M$ degrees of freedom  (spin, particle-hole pseudospin, transverse channel index, etc).
Accordingly, onsite terms $h_j$ and hopping terms $u_j$ are $M\times M$ matrices, while $\psi_j^{(\dag)}$ are column (row) vectors of fermion annihilation (creation) operators. 
We subdivide the system into a superconducting scattering region of $N$ sites ($j=1,...,N$), coupled to normal metallic homogeneous leads on the left and on the right.

To connect with optimal control, we consider an onsite Hamiltonian $h_j = h_j^{(0)} + \sum_k f_{kj} {\cal H}_k$, where $h_j^{(0)}$ is fixed and $k=1,\dots,p$ labels different tunable and spatially-varying parameters $f_{k j}$, such as the components of a magnetic field $\dB_j$, an electrostatic potential $V_j$ or a superconducting gap $\Delta_j$.
Our main goal is to perform an efficient numerical optimization of $f_{kj}$ in quantum wires.
For simplicity, we assume $u_j$ to be fixed and uniform, but our method can be generalized to relax this assumption, e.g. to optimize an inhomogeneous spin-orbit coupling.

\begin{figure}[t]
	\centering
	\includegraphics[]{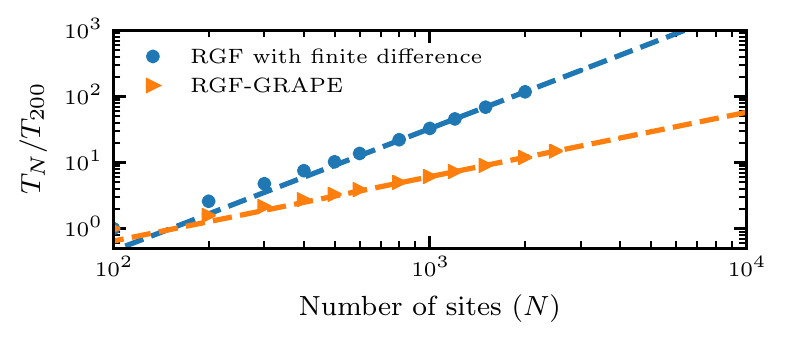}
	\caption{Power-law fits to the average computation time $T_N$ for the performance index (Eq.~\eqref{eq:phi_tilde}) and its gradient in a Majorana wire with $N$ sites. 
	The RGF-GRAPE method (orange triangles, $T_N\propto N^{0.98}$) is polynomially faster than the finite-difference gradient approach (blue disks, $T_N\propto N^{1.85}$). See appendix~\ref{subsec:bench} for details. 
	}
	\label{fig:benchmark}
\end{figure}
Similarly to the GRAPE algorithm, which iteratively constructs products of propagators to describe the system at each time step, local observables of a tight-binding lattice can be described in terms of propagators (Green's functions) obtained iteratively from the system's left and right boundaries.
This conceptual connection becomes concrete in the RGF method~\cite{Thouless:1981wq},
where the retarded Green's function at site $j$ and energy $E$ is written as
\begin{equation}
	G_{j}^{\mathrm{ret}}(E) = \parS{E- h_j - \Sigma_{j-1}^L - \Sigma^R_{j+1}}^{-1}.
	\label{eq:Gnn}
\end{equation}
Here, $\Sigma^{L(R)}_{j\pm1}$ is the left (right) hybridization function representing the influence of sites to the left (right) of site $j$.
These hybridization functions are obtained iteratively using the standard RGF recursion relations~(cf. Appendix~\ref{sec:rgf}). 

The recursive formalism shares two major advantages of GRAPE, in that it allows for a 
speedup of calculations through the reuse of intermediate results
 and it enables analytical expressions for the derivatives of propagators. 
As a result, the complexity of calculating a Green's function and its derivatives is reduced to $O(NM^3)$ (Fig.~\ref{fig:benchmark}).
In contrast, a naive finite difference approach for calculating $\partial G^{\rm ret}_j/\partial f_{k j'}$  for $j'=1,\dots,N$ would incur a total computational cost of  $O(N^2M^3)$, which can rapidly become prohibitive with the length of the system.

%%%%%%%%%%%%%%%%%%%%%%%%%%%%%%%%%%%%%%%%%%%%%%%%%%%%%%%%%%%%%%%%%%%%%%%%%%%%%%%
% Majorana functional and system considered 
%%%%%%%%%%%%%%%%%%%%%%%%%%%%%%%%%%%%%%%%%%%%%%%%%%%%%%%%%%%%%%%%%%%%%%%%%%%%%%%

\section{Majorana wire optimization}\label{sec:majOpt}
\subsection{Performance index definition}\label{ssec:index}
In order to optimize $\{f_{kj}\}$ for the realization of robust MBS, a performance index which is maximal for optimal spatial profiles is needed.
A good index must have the following attributes: 
(i) it is smooth under variations of $f_{kj}$; 
(ii) in the non-topological phase, the optimization process steers the system's parameters towards a topological phase transition (via gap closing); 
(iii) in the topological phase, the optimization evolves towards maximizing the protection of the MBS (via gap opening).
A simple performance index that meets the preceding criteria is 
\begin{equation}
	\Phi= -\Delta_L  Q_L  -\Delta_R Q_R,
	\label{eq:phi}
\end{equation}
where $\Delta_{L(R)}\geq 0$ is the local energy gap at the left (right) extremity of the scattering region and $Q_{L(R)}$ is the corresponding ``topological visibility''~\cite{DasSarma:2016ad}. 
For quantum wires belonging to symmetry class D~\cite{chiu2016},
the latter varies continuously between $\pm 1$ and its sign gives the $\mathbb{Z}_2$ topological invariant of the superconducting wire segment ($+/-1$ in the trivial/topological phase).

To benefit from the computational efficiency of the RGF method and the analogy to GRAPE, we express $\Phi$ in terms of Green's functions.
On the one hand, $Q_{L(R)}$ is given by the determinant of the zero-energy reflection matrix at site $j=0$ ($j=N+1$)~\cite{Akhmerov:2011pi,Fulga:2011qv,Fulga:2012kq}.
These matrices can be obtained~(cf. Appendix~\ref{sec:phiimp}) from $G_{0}^{\rm ret}(0)$ and $G_{N+1}^{\rm ret}(0)$ via the Fisher-Lee relations~\cite{Fisher:1981jt,Stone:1988lj,Baranger:1989hc,Wimmer:2008ve}. 
On the other hand, $\Delta_{L(R)}$ can be extracted from the spectral functions.
However, this requires evaluating $G_j^{\rm ret}(E)$ for multiple energies, which is numerically costly and inefficient. 
Fortunately, for the purposes of optimization, the absolute value of the gap is not needed, but only a function that scales in the same way.
Herein, we will construct an effective gap that is based solely on $G_j^{\rm ret}(0)$.

\begin{figure}[tb]
	\centering
        \includegraphics[width=0.95\linewidth]{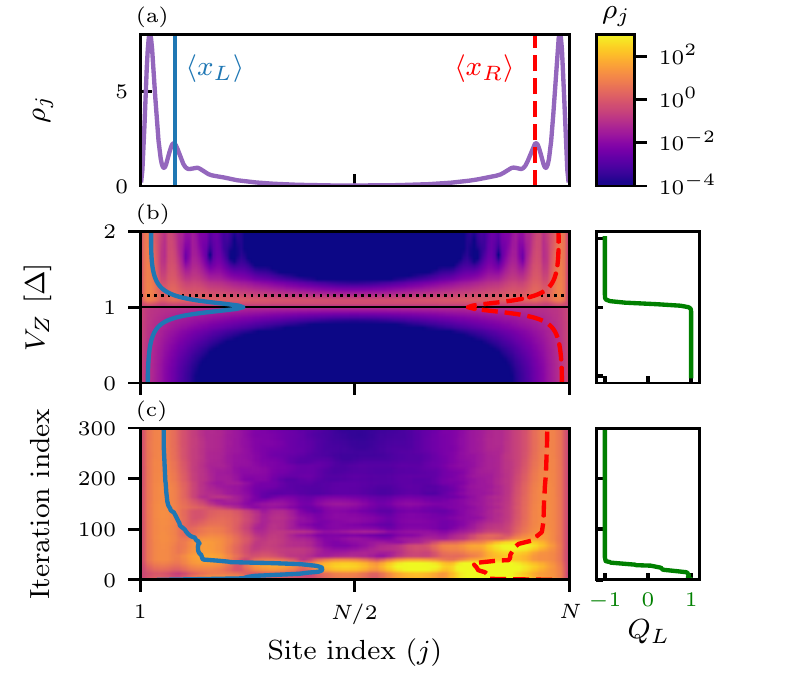}
	\caption{(a) Zero-energy local density of states $\rho_j$ and the corresponding ``centers-of-mass'' (CM) $\langle x_L\rangle$ and $\langle x_R\rangle$ (cf. Eq.~\eqref{eq:CM}) in a homogeneous wire with superconducting gap $\Delta$ and Zeeman splitting $V_Z$.
	(b) Evolution of $\rho_j$, the two CM, and the topological visibility $Q_L$ (green curve)  as a function of $V_Z$. 
	The black dotted line indicates the $V_Z$ value used in panel (a). 
	At the topological phase transition (solid black line), $\rho_j$ is approximately uniform.
	(c) Evolution of $\rho_j$, the CM and $Q_L$ with successive iterations of the RGF-GRAPE optimization algorithm. 
	}
	\label{fig1}
\end{figure}
%%%%%%%%%%%%%%%%%%%%%%%%%%%%%%%%%%%%%%%%%%%%%%%%%%%%%%%%%%%%%%%%%%%%%%%%%%%%%%%
% Heuristic gap
%%%%%%%%%%%%%%%%%%%%%%%%%%%%%%%%%%%%%%%%%%%%%%%%%%%%%%%%%%%%%%%%%%%%%%%%%%%%%%%

To that end, we define the ``\emph{center-of-mass}'' (CM) of the left (L) and right (R) zero-energy states (Fig.~\ref{fig1}a),
\begin{align}
\label{eq:CM}
\langle x_L \rangle &= \frac{ 1 }{\Nm_L}\sum_{j=1}^{N/2} j \rho_j\,\,;\,\,  
\langle x_R \rangle = \frac{ 1 }{\Nm_R}\sum_{j=N/2}^{N} j \rho_j, 
\end{align}
where $\rho_j \equiv -\im \parC{\tr \parS{ G_j^{\ret} (0)}}/(2\pi)$ is the zero-energy local density of states, while $\Nm_L = \sum_{j=1}^{N/2} \rho_j$ and $\Nm_R = \sum_{j=N/2}^{N} \rho_j$ are the total ``masses'' of the L and R states.
In a superconducting wire weakly coupled to normal leads, zero-energy states from the leads leak in the superconducting region leading to  $\rho_j\neq 0$ in both the topological and trivial phase. 
Hence, the CM gives a smooth and quantitative measure of the localization of zero-energy states  (Fig.~\ref{fig1}b).
Moreover, there is an inverse relation between the localization length of $E=0$ states and the $p-$wave component of the superconducting gap~\cite{Potter:2011sr}: the larger the latter is, the closer $\langle x_L\rangle$ and $\langle x_R\rangle$ get to $1$ and $N$, respectively.
With this in mind, we introduce the effective gaps 
\begin{align}
	\tilde{\Delta}_L &\equiv \frac{N/2}{\langle x_L \rangle - 1 }\,\,;\,\,
	\tilde{\Delta}_R\equiv \frac{N/2}{N - \langle x_R \rangle },
\label{eq:h_gap}
\end{align}
which lead to a performance index
\begin{equation}
\label{eq:phi_tilde}
	\tilde{\Phi} = -\tilde{\Delta}_L  Q_L  -\tilde{\Delta}_R Q_R.
\end{equation}
The effective gaps $\tilde{\Delta}_{L(R)}$ correlate closely with the $p-$wave component of the superconducting gap. 
The optimization of $\tilde{\Phi}$ will accordingly converge towards maximally localized MBS, which is beneficial for the phase coherence of MBS-based qubits~\cite{Knapp:2018}.
However,  $\tilde{\Delta}_{L(R)}$ may be blind to localized non-topological subgap states that might appear at the extremities of the wire, 
insofar as these 
do not affect the localization length of the MBS.
Though these subgap states can lead to a ``soft'' gap, their occupation does not flip the MBS-based qubit's parity if the MBS are spatially well-separated~\cite{Potter:2011sr,Akhmerov:2010kl}.

%%%%%%%%%%%%%%%%%%%%%%%%%%%%%%%%%%%%%%%%%%%%%%%%%%%%%%%%%%%%%%%%%%%%%%%%%%%%%%%
% Reality checks
%%%%%%%%%%%%%%%%%%%%%%%%%%%%%%%%%%%%%%%%%%%%%%%%%%%%%%%%%%%%%%%%%%%%%%%%%%%%%%%
\subsection{Proof of concept of the optimization}\label{ssec:proof}
Our optimization algorithm, which we name RGF-GRAPE, consists of the following steps: 
(i) propose an initial set of $\{f_{kj}\}$;
(ii) compute $\tilde{\Phi}$ and $\partial\tilde{\Phi}/\partial f_{kj}$ adapting ideas from RGF and GRAPE~(cf. Appendix~\ref{sec:phiimp});
(iii) update $\{f_{kj}\}$ via gradient ascent;
(iv) repeat steps (ii) and (iii) until a maximum of $\tilde{\Phi}$ is attained.
These steps are applicable to an arbitrary single-particle tight-binding model of nanowire.
The occurence of soft gaps can be reduced by running the algorithm with varying initial and optimization parameters and keeping the solution with the largest gap afterwards (cf. Appendices~\ref{regularization} and \ref{convergence}).

To confirm that the algorithm is working properly, 
we consider a superconducting wire in the single channel regime without spin-orbit coupling placed under an inhomogeneous magnetic field. 
Starting from a topologically trivial initial state (a superposition of two spiraling fields of different periods), successive iterations of the optimization algorithm adjust the magnetic texture and drive the wire into the topological regime (Fig.~\ref{fig1}c).
The final magnetic texture, shown in Appendix~\ref{boundaries}, resembles a perfect spiral in the bulk of the wire, but departs from it within a superconducting coherence length from the boundaries.
At a small loss for the topological gap,
the departure from a uniform spiral renders the MBS significantly more localized, which lead to a MBS energy splitting reduced by more than an order of magnitude.
This finding demonstrates that boundary engineering, appropriately done, can improve the MBS characteristics. 
It also shows that in inhomogeneous wires, unlike in uniform ones,  the zero-mode energy splitting can be suppressed without increasing the topological gap or the length of the wire.

%%%%%%%%%%%%%%%%%%%%%%%%%%%%%%%%%%%%%%%%%%%%%%%%%%%%%%%%%%%%%%%%%%%%%%%%%%%%%%%
% Electrostatic gate optimization
%%%%%%%%%%%%%%%%%%%%%%%%%%%%%%%%%%%%%%%%%%%%%%%%%%%%%%%%%%%%%%%%%%%%%%%%%%%%%%%
\begin{figure}[tb]
	\centering
	\includegraphics[width=\linewidth]{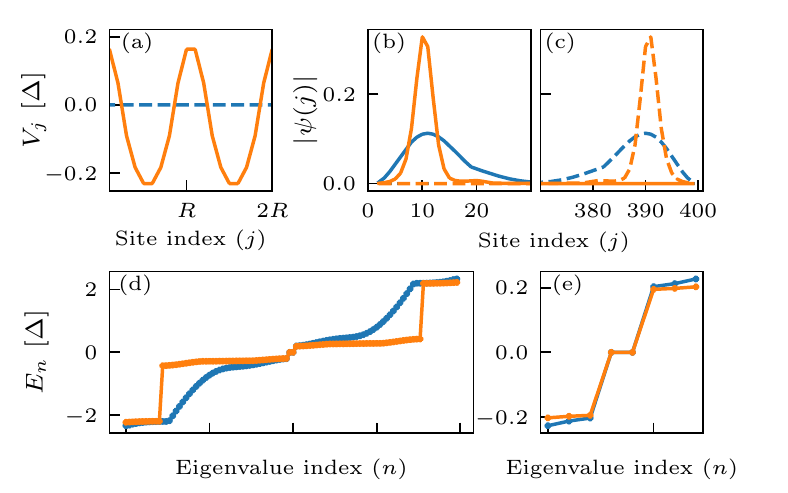}
	\caption{Optimization of the electrostatic 
          potential profile
          in a 
	  spin-orbit coupled superconducting wire with $400$ sites and one subband.
          Blue (orange) curves are the results before (after) optimization. 
	  (a) Initial and final (optimized) gate potential textures. A period of $R=10$ sites (lattice constant $a=10$~nm) is enforced during the optimization.
	 (b,c) MBS wave function amplitude.
         (d) Energy eigenvalues ($E_n$) of the isolated superconducting region. 
	(e) Energy spectrum near zero-energy. See Appendix~\ref{sec:modelAndParams} for parameters.
	}
	\label{mu_modulation}
\end{figure}
\subsection{Optimization of potential profiles}\label{ssec:gate}
The electrostatic potential in realistic quantum wires is inevitably inhomogeneous and partially tunable by applying voltages on nearby gates.
Recent studies~\cite{Prada:2012, Kells:2012, Bagrets:2012,  Moore:2016} have concluded that inhomogeneities result in non-topological localized states with near-zero energy. 
Other authors~\cite{DeGottardi:2013a} have analyzed the impact of (quasi)periodic gate potentials in the topological phase diagram.
Yet, there are no explicit results about the optimal spatial profile that would lead to more robust MBS.

In Fig.~\ref{mu_modulation}, we perform an optimization of the electrostatic potential profile.
The precise relation between this potential and the gate voltage could be obtained by solving the Schrodinger-Poisson equation~\cite{Vuik:2016wc,Woods2018,Antipov2018,Mikkelsen2018}.
For simplicity, we constrain ourselves to smooth and periodic potentials along the wire axis.
Non-periodic potentials are non-optimal in that they generically lead to soft gaps~\cite{Moore:2016}.
Smoothness can be achieved by imposing penalties in $\tilde{\Phi}$ against rapid potential variations (cf. Appendix~\ref{regularization}). 
Through successive iterations of the optimization algorithm, the potential profile evolves from a uniform initial state to a harmonic final state.
Unexpectedly, harmonic modulations strongly enhance the MBS localization, while preserving the initial energy gap (albeit with a larger density of states at the gap edge).
The increased localization reduces the MBS energy splitting by three orders of magnitude. 
Such improvement could be crucial for extending coherence times in Majorana-based qubits, whose dephasing times are expected to be limited by the zero-mode energy splittings~\cite{Knapp:2018}. This result appears to be a counterexample to the common belief that inhomogeneous potentials are harmful for MBS.

\begin{figure}[tb]
	\centering
	\includegraphics[width=\linewidth]{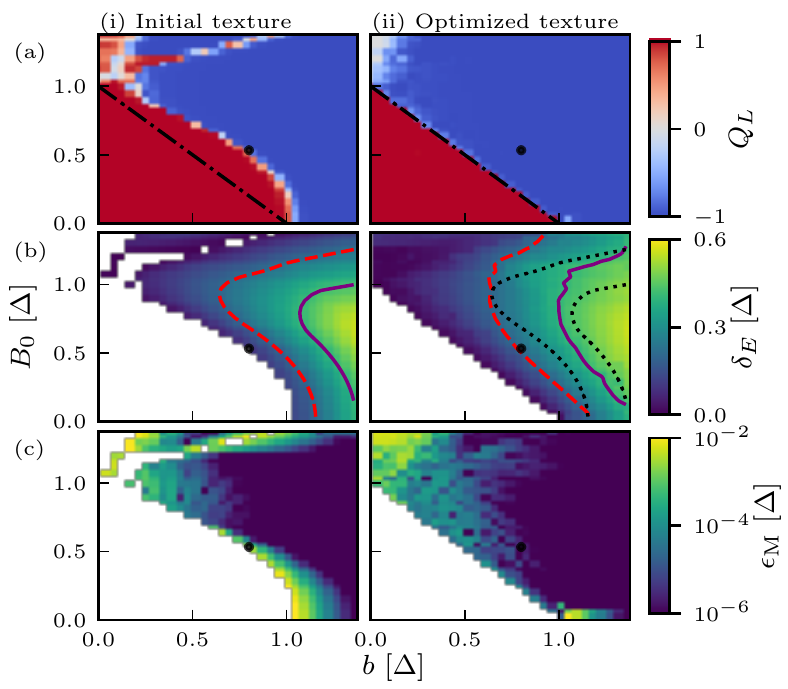}
	\caption{Characteristics of Majorana wires with experimentally relevant parameters~(cf. Appendix~\ref{sec:modelAndParams}) with (i) non-optimized magnetic texture (combination of uniform and spiral fields with period $R=N/16=25$) and (ii) optimized magnetic texture in the $b$-$B_0$ parameter space. 
	For each point in parameter space, we take the outcome with the largest energy gap out of multiple optimizations with different optimization parameters (cf. Appendix~\ref{regularization}).
	The same optimized texture is used for all three rows.
	(a) Topological visibility $Q_L$ (red: trivial phase, blue: topological phase). The black dot-dashed line indicates $B_0 + b = \Delta$, the expected minimal Zeeman energy to reach the topological phase.
	(b) Energy gap $\delta_E$ in the topological phase (white regions indicate the trivial phase).
	Dotted black curves are the contours drawn in the panel to the left. Dashed (solid) curves denote energy gaps of 100~(200)~mK. 
	(c) Majorana zero-mode energy splitting $\epsilon_M$ in the topological phase.
	}
	\label{fig:phaseSpace}
\end{figure}
%%%%%%%%%%%%%%%%%%%%%%%%%%%%%%%%%%%%%%%%%%%%%%%%%%%%%%%%%%%%%%%%%%%%%%%%%%%%%%%
% Superconducting gap
%%%%%%%%%%%%%%%%%%%%%%%%%%%%%%%%%%%%%%%%%%%%%%%%%%%%%%%%%%%%%%%%%%%%%%%%%%%%%%%
In view of the preceding result, one might question whether a spatially uniform superconducting gap is optimal or not. 
According to RGF-GRAPE, the answer turns out to be affirmative (not shown), this time in agreement with conventional wisdom~\cite{Takei:2013,Lutchyn2017}.

%%%%%%%%%%%%%%%%%%%%%%%%%%%%%%%%%%%%%%%%%%%%%%%%%%%%%%%%%%%%%%%%%%%%%%%%%%%%%%%
% Result: Majorana in new parameter regime
%%%%%%%%%%%%%%%%%%%%%%%%%%%%%%%%%%%%%%%%%%%%%%%%%%%%%%%%%%%%%%%%%%%%%%%%%%%%%%%
\begin{figure}[t]
	\centering
	\includegraphics[width=\linewidth]{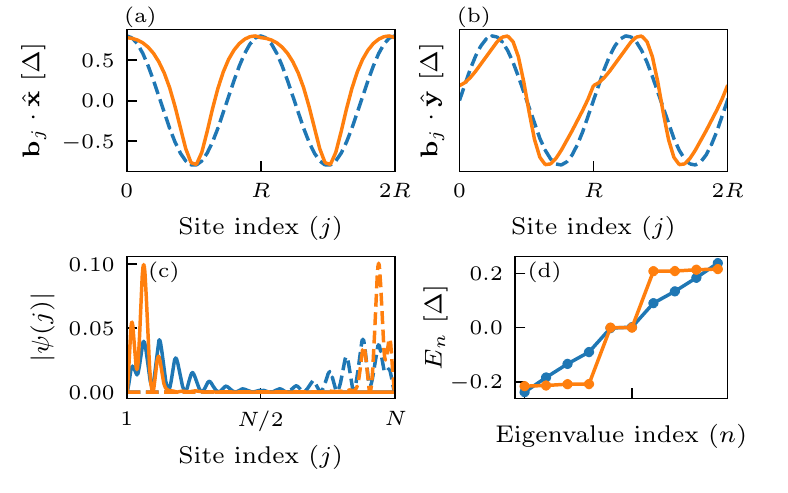}
	\caption{Optimized magnetic textures for $B_0 = 0.53 \Delta$ and $b = 0.8 \Delta$ (black disks of Fig.~\ref{fig:phaseSpace}).
	Blue (orange) curves correspond to initial (optimized) textures.
	(a,b) Components of the magnetic textures ($R=N/16 = 25$).
	(c) MBS wave function amplitude. 
	(d) Energy spectrum near zero energy. 
	}
	\label{fig:phaseSpaceExample}
\end{figure}
\subsection{Optimization of magnetic textures}\label{ssec:texture}
Inhomogeneous magnetic fields produced by arrays of micromagnets constitute a tunable resource for the emergence and manipulation of MBS~\cite{Fatin:2016}.
On the one hand, spiral fields lead to an artificial spin-orbit coupling that can induce topological superconductivity in weakly spin-orbit coupled wires~\cite{Kjaergaard:2012tg}.
On the other hand, the combination of spiral and uniform magnetic fields can help attain MBS when neither of the fields alone would suffice~\cite{Klinovaja:2012}.
However, once again little is known about the optimal magnetic texture conducive to more robust MBS.
The RGF-GRAPE algorithm is well suited to explore this issue.

We consider a superconducting wire without intrinsic spin-orbit coupling, subjected to a uniform magnetic field ${\bf B}_0$ and a spiral-like magnetic field ${\bf b}_j$. 
We assume the amplitude of ${\bf b}_j$ to be uniform and equal to $b$, while its site-dependent orientation is optimized using RGF-GRAPE.
Figure~\ref{fig:phaseSpace} compares key attributes of Majorana wires between the case where ${\bf b}_j$ is a perfect spiral (column (i)) and the case where ${\bf b}_j$, along with a uniform chemical potential, are optimized (column (ii)). 
From the topological visibility in panel (a), one can see that the optimization reaches the topological phase as long as $B_0+b >\Delta$.
In addition, the constant topological gap contours in panel (b) show that the optimization allows to increase the parameter space area where the gap is larger than experimentally relevant temperatures.
It is likewise clear that, for a fixed $b$, adding a modest uniform field augments the topological gap.
This result offers a path to circumventing the problem of small g-factors, which has impeded the experimental realization of MBS in wires with weak intrinsic spin-orbit coupling.
Finally, from the zero-mode energy splitting in panel (c), it ensues that the optimization allows to greatly enhance the zero-mode degeneracy, in particular in the low $B_0$ region.
These findings are useful for the realization of MBS using micromagnet arrays~\cite{Maurer:2018,Kjaergaard:2012tg}, where $b_j$ is limited to~$\lesssim 1$~T.

Figure~\ref{fig:phaseSpaceExample} gives a more detailed account of the optimization for a fixed amplitude of $B_0$ and $b$ (the black disks in Fig.~\ref{fig:phaseSpace}).
In this case, the optimized texture results in a large enhancement of the topological gap (from less than $50$ mK to $100$ mK) and a large reduction of the MBS' localization length that suppresses the MBS energy-splitting by more than two orders of magnitude. 
This finding suggests that small but judicious departures from simple textures can significantly improve the MBS attributes.

%%%%%%%%%%%%%%%%%%%%%%%%%%%%%%%%%%%%%%%%%%%%%%%%%%%%%%%%%%%%%%%%%%%%%%%%%%%%%%%
% Outlook and conclusion
%%%%%%%%%%%%%%%%%%%%%%%%%%%%%%%%%%%%%%%%%%%%%%%%%%%%%%%%%%%%%%%%%%%%%%%%%%%%%%%
\section{Conclusion}\label{sec:conclusion}
We have introduced an efficient algorithm that optimizes real-space parameter profiles in superconducting quantum wires for the generation of robust Majorana bound states. 
The algorithm explores regions of parameter space where no intuitive (analytical) results are available and identifies new regimes for the emergence of MBS with strongly reduced energy splitting.
Combined with realistic device simulations,
 our algorithm could provide detailed guidance for improved coherence in Majorana-based qubits.
More generally, variations of the introduced RGF-GRAPE algorithm could be applied to characterize new topological phases in inhomogeneous low dimensional systems 
including photonic crystals where machine learning has recently been used for a similar purpose~\cite{Pilozzi2018}.

\begin{acknowledgments}
This work was funded by the Canada First Research Excellence Fund and by the National Science and Engineering Research Council. 
Numerical calculations were done with computer resources from Calcul Québec and Compute Canada.
The authors benefited from fruitful discussions with P. Lopes,  M. Pioro-Ladrière and S. Turcotte.
The source code for this work is available at~\footnote{
S. Boutin, J. Camirand Lemyre and I. Garate, \href{http://dx.doi.org/10.5281/zenodo.1486048}{"Majorana bound state engineering via efficient real-space parameter optimization,"} (2018), source code.
[http://doi.org/10.5281/zenodo.1486048]
}.
\end{acknowledgments}

\appendix 

\section{Gradient Ascent Pulse Engineering}\label{sec:grape}
This appendix is a short self-contained introduction to the GRAPE algorithm~\cite{Khaneja:2005tw}, with a focus on its computational complexity. The goal is to make explicit the analogy between the GRAPE algorithm, the RGF method, and the RGF-based optimization algorithm introduced in Sec.~\ref{sec:realspaceOCT}.

For the sake of simplicity, we follow Ref.~\cite{Rowland:2012vy} and present the GRAPE algorithm for a specific optimal control problem: the optimization of a time-dependent Hamiltonian for the preparation of a target unitary transformation $V$ in a time $T$. Given a time-dependent Hamiltonian
\begin{equation}
	H(t) = H_0  + \sum_{k=1}^p f_k(t) \mathcal{H}_k,
\end{equation}
this control problem can be stated as finding the $p$ control functions $\parC{f_k(t)}$ such that the propagator resulting from time-evolution, $U(T)$,
realizes the target transformation $V$. 
One can quantify the success of a solution using the performance index 
$\Phi = \tr \parS{V^\dag U(T)}$, which is the inner product between the realized and the target propagators~\footnote{
	As $\Phi$ is in general a complex number, the actual performance index should be $|\Phi|^2$~\cite{Rowland:2012vy}. While important in practice, this distinction does not influence the description of the algorithm and the analysis of its computational complexity. 
}.
One can find a solution to the control problem by maximizing this performance index. 
While optimization algorithms, such as gradient descent, are commonplace and independent of the problem, the GRAPE algorithm uses knowledge of the structure of the propagator to calculate efficiently the gradient of $\Phi$, which can then be used by an optimization algorithm.

In general, the propagator $U(T)$ is complicated to calculate, as it involves a time-ordered exponential. However, the problem can be greatly simplified by considering the functions $f_k(t)$ as piecewise constant functions~\cite{Khaneja:2005tw}. In this reduced optimization space, the propagator is
\begin{equation}
	U(T) =
	U_N U_{N-1} \cdots U_2U_1,
\end{equation}
where, for timestep $t_j$ of duration $\Delta t = T/N$, the propagator of the locally time-independent Hamiltonian is
\begin{equation}
	U_j = \exp \parC{ 
	- i \Delta t \, \parS{H_0 + \sum_{k=1}^p f_k(t_j) \mathcal{H}_k }
	 }.
\end{equation}
By removing the time-ordering operator from the problem, the derivatives of the performance index with respect to the (now finite) set of control parameters $\parC{ f_k(t_j) }$ can be easily obtained using the linearity of the trace
\begin{align}
	\frac{ \del \Phi }{\del f_k(t_j)} &= 
	\tr \parS{ V^\dag U_N \cdots U_{j+1} 
	\frac{\del U_j  }{\del f_k(t_j)} U_{j-1} \cdots U_1 }
	\\
	&=
	\tr \parS{ P_{j}^\dag
	\frac{\del U_j  }{\del f_k(t_j)} X_{j-1}},
\end{align}
where in the second line we have defined the forward-in-time string of propagators $X_j = U_jU_{j-1}\cdots U_1$ and the backward-in-time string of propagators $P_{j} = U_{j+1}^\dag \cdots U_N^\dag V$.
These strings of propagators are at the origin of the computational advantage of the GRAPE algorithm over more naive finite-difference approaches. In brief, 
due to the recursive nature of these strings (e.g. $X_{j+1} = U_{j+1} X_j$ and $P_{j-1} = U_j^\dag P_{j}$), the computational cost of computing the final strings $X_N$ and $P_1$  is the same as computing all the strings if one simply keeps intermediate results in memory. Thus, for the computational cost of a single forward-in-time evolution (calculation of $X_N$) and a single backward-in-time evolution (calculation of $P_1$) one can compute the performance index and all of its derivatives.

For concreteness, we summarize the GRAPE algorithm indicating, where appropriate, the computational cost of the step in brackets:
\begin{enumerate}\setlength{\itemsep}{0pt}
	\item Choose initial vector of parameters $\parC{ f_k(t_j) }$.
	\item Calculate and store in memory all propagators $U_j$ \newline
	[$N$ matrix exponentials].
	\item Starting from $X_0=\id$, calculate  and store in memory all forward propagator strings [$N$ matrix products].
	\item Starting from $P_N = V$ calculate and store in memory  all backward propagator strings [$N$ matrix products].
	\item Calculate gradient of $\Phi$ with respect to $\parC{ f_k(t_j) }$ [$2pN$ matrix products].
	\item Use gradient to update the parameter vector and return to step 2.
\end{enumerate}
Steps 2-5 are the core of the GRAPE algorithm, while step 1 and 6 are general steps of any gradient-based optimization algorithm. One can see that all steps are at most linear in $N$. 
If one does not keep intermediate results in memory and computes each derivatives independently (as is usually the case in a finite-difference calculation), the complexity is $O(N^2)$. Thus, at the cost of an increased usage of memory, the GRAPE algorithm allows a polynomial speedup (in the number of timesteps) over a finite-difference approach.

%%%%%%%%%%%%%%%%%%%%%%%%%%%%%%%%%%%%%%%%%%%%%%%%%%%%%%%%%%%%%%%%%%%%%%%%%%%%%%%
% SEC II RGF-based opt
%%%%%%%%%%%%%%%%%%%%%%%%%%%%%%%%%%%%%%%%%%%%%%%%%%%%%%%%%%%%%%%%%%%%%%%%%%%%%%%
\section{RGF-based optimization}\label{sec:rgf}
In this appendix, we give more details about the real-space optimization algorithm (which we name RGF-GRAPE) based on the RGF method. After stating the useful recursive relations and their derivatives, we expand on its relation to the GRAPE algorithm and its computational complexity. 

\subsection{Recursive relations}
Following the notation set in Eq.~\eqref{eq:hTB}, we consider a 1D tight-binding Hamiltonian with onsite terms $h_j$ and nearest-neighbor hopping terms $u_j$.
We assume there are M local degrees of freedom per site.
As stated in Eq.~\eqref{eq:Gnn},
the RGF method~\cite{Thouless:1981wq} allows to write the retarded Green's function of the system, projected onto site $j$,
as a function of the hybridization function $\Sigma_{j\pm1}^{L(R)}$ which describres the influence on site $j$ of the sites to the left (right).
These hybridization functions are defined as
\begin{equation}
	\Sigma^{L}_{j} = u_j G^L_j u_j^\dag;
	\qquad
	\Sigma^{R}_{j} = u_{j-1}^\dag G^R_j u_{j-1},
	\label{eq:SlSr}
\end{equation}
where $G^{L(R)}_j$ is the projection on site $j$ of the Green's function of the system formed by site $j$ and all sites to its left (right).
The exact expressions for these Green’s functions are obtained iteratively by using the following recursion relations (see e.g. Refs.~\cite{Wimmer:2008ve, Lewenkopf:2013kk} for reviews):
\begin{align}
	\begin{split}
	G^L_j &= (E- h_{j} - \Sigma_{j-1}^{L} )^{-1}
		\\
		G^R_j &= (E- h_{j} - \Sigma_{j+1}^{R} )^{-1}.\end{split}
	\label{eq:GlGr}
\end{align}
The initial left (right) lead hybridization function $\Sigma^{L(R)}_{0(N+1)}$, starting the recursion relation, can be calculated using the translation invariance of the lead defined by the sites $j\leq 0$ ($j\geq N+1$). Indeed, the translation symmetry in the semi-infinite leads implies the relations $\Sigma^L_{j\leq0} = \Sigma^L_0$ and $\Sigma^R_{j>N} = \Sigma^R_{N+1}$.
In our numerics, these lead hybridization functions are calculated using the \textsc{Kwant} numerical package~\cite{Groth:2014qc}.

Before considering derivatives of Green's functions and their use for efficient optimization, it is worth noting that, by itself, the RGF method has many similarities to the GRAPE algorithm. 
Indeed, both methods keep in memory intermediary results of recursive relations relating propagators to reduce computational complexity.
The steps of the RGF algorithm for computing all lattice Green's functions $G_j^\ret$, including complexity in brackets, are: 
\begin{enumerate}\setlength{\itemsep}{0pt}
	\item Compute $\Sigma^L_0$ and $\Sigma^R_{N+1}$ [Independent of $N$: $O(M^3)$].
	\item Starting from $\Sigma^L_0$ compute and store in memory all $\Sigma^L_j$ [$N$ matrix inversions, $2N$ matrix products: $O(NM^3)$].
	\item Starting from $\Sigma^R_{N+1}$ compute and store in memory  all $\Sigma^R_j$ [$N$ matrix inversions, $2N$ matrix products: $O(NM^3)$].
	\item Calculate all $G_j^{\ret}$ [$N$ matrix inversions: $O(NM^3)$].
\end{enumerate}
Steps 2-4 of this algorithm are analog to steps 3-5 of the GRAPE algorithm stated in Sec.~\ref{sec:grape}, where the time-domain propagators have been replaced by real-space lattice Green's functions.
To make even clearer the analogy between the RGF method and GRAPE, one can restate the recursive relations of Eq.~\eqref{eq:GlGr} as a string of enlarged matrix products using properties of the so-called Mobiüs transformation~\cite{Umerski:1997tg}.

Finally, from a computational complexity point-of-view, by keeping in memory all the hybridization functions $\Sigma^L_j$, $\Sigma^R_j$, one can compute with complexity $O(NM^3)$ all onsite lattice Green's functions. This is a polynomial speedup over a naive inversion of the full Hamiltonian, which is an $O(N^3M^3)$ calculation. Such a speedup is possible due the nearest-neighbor hopping structure of the problem, which leads to a block-tridiagonal matrix representation of the single-particle Hamiltonian.

\subsection{Derivatives of the RGF expressions}
Taking the onsite Hamiltonian to be $h_j = h_j^{(0)} + \lambda_j A$, with $A$ some local operator~\footnote{
Although we consider here a site-independent operator, our calculation naturally extends to site-dependent local operators with $A \ra A_j$.
},
we now calculate the derivative of the lattice Green's function at site $j$ with respect to a local parameter of a possibly different site $\lambda_n$. 
Using standard matrix algebra, this derivative is
\begin{align}
	\begin{split}&
	\frac{ \del }{\del \lambda_n} G^\ret_{j}(E) = 
	% \\&
		G_{j}^{\mathrm{ret}}
			\parS{ \delta_{n,j} A + 
			\frac{ \del }{\del \lambda_n}
			\parO{\Sigma_{j-1}^L 
						+
						 \Sigma_{j+1}^R}}
		G_{j}^{\mathrm{ret}},
		\end{split}
	\label{eq:dGRet}
\end{align}
which can be expanded using the definition of the left and right hybridization functions. These derivatives are given by
\begin{align}
	\frac{ \del }{\del \lambda_n}
		\Sigma_{j}^L 
		&= u_j  G^L_j \parS{ \delta_{n,j}A + \frac{ \del }{\del \lambda_n} \Sigma^L_{j-1} } G^L_j u_j^\dag;
	\\
	\frac{ \del }{\del \lambda_n}
		\Sigma_{j}^R 
		&= u_{j-1}^\dag  G^R_j \parS{ \delta_{n,j}A + \frac{ \del }{\del \lambda_n} \Sigma^R_{j+1} } G^R_j u_{j-1}.
\end{align}
In order to implement these expressions in a computer program, it is useful to rewrite them as 
\begin{align}
	\frac{ \del }{\del \lambda_n} \Sigma^L_j &= 
			\Theta(j-n) % n is to the left 
			\parO{u_j G^L_j}  \parO{u_{j-1}G^L_{j- 1}} \cdots
			\\ & \nonumber
			 \times \parO{u_nG^L_n} 
			A
			\parO{G^L_n u_n^\dag }
			\cdots \parO{G^L_{j- 1}u_{j-1}^\dag} 
			\parO{G^L_j u_j^\dag},
	\\
	\frac{ \del }{\del \lambda_n} \Sigma^R_j &= 
		\Theta(n-j) % n is to the right
		\parO{u^\dag_{j-1} G^R_j}  
		\parO{u_{j}^\dag G^R_{j+1}} \cdots 
		\\ & \nonumber \times
		\parO{u_{n-1}^\dag G^R_n} 
		A
		\parO{G^R_n u_{n-1} }
		\cdots \parO{G^R_{j- 1}u_{j-2}} 
		\parO{G^R_j u_{j-1}},
\end{align}
where $\Theta$ is the Heaviside function with $\Theta(j\geq0)=1$.
By analogy to the GRAPE algorithm, we now define strings of propagators, such that derivatives can be simply expressed as 
\begin{equation}
	\begin{split}
	\frac{ \del }{\del \lambda_n} \Sigma^L_j =
			\Theta(j-n) % n is to the left  
			X^L_{j,n}\frac{ \del h_n}{\del \lambda_n} 
			P^L_{j,n};
		% \qquad
		\\
		\frac{ \del }{\del \lambda_n} \Sigma^R_j = 
		\Theta(n-j) % n is to the right
			X^R_{j,n}\frac{ \del h_n}{\del \lambda_n} 
			P^R_{j,n},\end{split}
		\label{eq:dSXP}
\end{equation}
where $\frac{ \del h_n }{\del \lambda_n} =A$
and the Heaviside function is used to make explicit that, by definition, a left (right) hybridization function can not have a nonzero derivative with respect to a parameter to its right (left).
The explicit definitions of the propagator strings in recursive form are
\begin{align}
	\begin{split}
	X^L_{j,n} &= 
			\parO{u_j G^L_j}  \parO{u_{j-1}G^L_{j- 1}} \cdots \parO{u_nG^L_n}
			\\ &= X^L_{j,n+1} \parO{u_nG^L_n}
		\\
		P^L_{j,n} &= 
			\parO{G^L_n u_n^\dag } \cdots \parO{G^L_{j- 1}u_{j-1}^\dag} 
			\parO{G^L_j u_j^\dag}
			\\&= \parO{G^L_n u_n^\dag } P^{L}_{j, n+1}
			% ;\nonumber
		\\
		X^R_{j,n} &= 
			\parO{u^\dag_{j-1} G^R_j}  
			\parO{u_{j}^\dag G^R_{j+1}} \cdots \parO{u_{n-1}^\dag G^R_n}
			\\&=
			X^R_{j,n-1} \parO{u_{n-1}^\dag G^R_n}
			% ;\nonumber
		\\
		P^R_{j,n} &= 
		\parO{G^R_n u_{n-1} }
			\cdots \parO{G^R_{j- 1}u_{j-2}} 
			\parO{G^R_j u_{j-1}}
			\\&=
			\parO{G^R_n u_{n-1} } P^{R}_{j,n-1}
			.\end{split}
		\label{eq:rgfStrings}
\end{align}
Similar recursive definitions can also be written for the $j$ index. 

\subsection{RGF-based real-space optimization} \label{subsec:RGFopt}
Using the results of the previous subsections, one can now build an algorithm similar to GRAPE for the calculation of the derivative of lattice Green's functions with respect to the real-space profile of parameters.
To make the analogy to GRAPE clearer, we first consider a performance index which depends only on a single Green’s function.
For example, when considering the Majorana wire optimization, 
this would be the case if close to the topological phase transition one considers directly the topological visibility as a performance index such that  $\Phi = -\det(r) = -F \parS{ G_{0}^\ret }$, with  $G_{0}^\ret$ the Green's function projected on site 0, and $F$ a function defined in App.~\ref{sec:Fisher}. 

The RGF-GRAPE optimization algorithm can be summarized in a way very similar to the GRAPE algorithm. As in the previous sections, we state the main steps of the algorithm and their respective computational complexity:
\begin{enumerate}\setlength{\itemsep}{0pt}
	\item Choose initial vector of parameters $\parC{ f_{k,j}}$.
	\item Calculate $G^\ret_j$ using the RGF method and storing all $G^{L(R)}_j$ in memory [$O(NM^3)$].
	\item Starting from $X^L_{j,j}=u_jG^L_j$, calculate $X^L_{j,1}$ storing intermediate results in memory ($X^L_{j,j-1}$, $X^L_{j,j-2}$,\dots). Similarly compute the strings of propagators for $P^L_{j,n}$, $X^R_{j,n}$, and $P^R_{j,n}$.\newline
	[Calculating all strings requires $4N$ matrix products: $O(NM^3)$].
	\item Compute derivatives using Eqs.~\eqref{eq:dSXP} and~\eqref{eq:dGRet}.
	\item Use gradient to update the parameter vector and restart to step 2.
\end{enumerate}
Comparing the GRAPE algorithm stated in App.~\ref{sec:grape}, one can see that the structure of the gradient calculation performed in steps 2-4 is very similar. This similarity extends to the computational complexity, such that the derivative of a lattice Green's function at a given site $j$ (fixed) with respect to parameters on each site ($n=1,\dots N$) scales linearly with the number of sites in the scattering region. More precisely, it is the same as the RGF calculation: $O(NM^3)$. In the case of a finite-difference calculation, where one would perform the RGF calculation $N$ times in order to vary each parameter to be optimized, the complexity would be $O(N^2M^3)$. Thus, by using the above algorithm, one obtains a polynomial speedup over finite differences.

If we now extend the above algorithm to a more general performance index, which requires the derivative of lattice Green's functions at all sites, the complexity becomes $O(N^2M^3)$ (i.e., the same as for finite-difference). 
This is a consequence of the fact that, in that most general case, we need to vary both indices of the propagator strings defined in Eq.~\eqref{eq:rgfStrings}, which  requires more matrix products. In the GRAPE analogy, this would be equivalent to having a performance index that depends on the propagator at multiple times.

Below, we consider in more detail the case of the LDOS-based performance index defined in Sec.~\ref{ssec:index} for the study of Majorana wires. 
This index depends on multiple Green's functions belonging to different sites. Nevertheless, by exploiting the structure of the performance index, the computational complexity of the gradient calculation can be made linear in $N$.

\section{Performance index implementation}\label{sec:phiimp}
In this appendix, we expand on the implementation of the LDOS-based performance index for Majorana wire optimization defined in Eq.~\eqref{eq:phi_tilde}. 
We first state the Fisher-Lee relations used to relate the calculation of the topological visibility to lattice Green's functions.
Then, we discuss how to efficiently implement the gradient of the effective gaps defined in Eq.~\eqref{eq:h_gap}.
Finally, we verify numerically the complexity of various gradient calculations.

\subsection{Fisher-Lee relations and the topological visibility}\label{sec:Fisher}
The scattering matrix can be obtained from the Green's function using the Fisher-Lee relations generalized to account for the presence of a magnetic field~\cite{Baranger:1989hc}.
Following the notation of Ref.~\cite{Wimmer:2008ve}, in the case of a scattering region of $N$ sites (site index $j = 1, \dots, N$) connected to a lead to the left (L) at site $0$ and to a lead to the right (R) at site $N+1$, the matrix elements of
the left reflection matrix $r^{L}$ are given by 
\begin{align}
	r_{mn}^{L} &= 
	   \tilde{\phi}^{L \dag}_{m,\mathrm{out}} \Gamma_{L} 
	   \parS{ iG^{\ret}_{0} \Gamma_L  - \id}\tilde{\phi}^{L}_{n, \mathrm{in}},
	   \label{eq:FisherLee}
\end{align}
with $\tilde{\phi}^{(l)}_{m, \alpha}  = \phi^{(l)}_{m, \alpha} / \sqrt{\hbar v_{m, \alpha}}$ the current normalized wavefunctions of propagating modes $m$ in lead $l$ with mode velocity $v_{m,\alpha}$, and $\Gamma_L = i \parS{\Sigma_0^L - (\Sigma_0^L)^\dag}$ the skew-hermitian part of the surface self-energy of the first site of the left lead. A similar expression for $r^{R}_{mn}$ can be obtained under the index changes $L\leftrightarrow R$ and $0 \leftrightarrow N+1$. To lighten the notation, the energy dependence of $\phi_{m,\alpha}^{L}$, $\Gamma_L$ and $G_{0}^{\ret}$ has been suppressed. 

The calculation of the Majorana wire performance index as defined in both Eq.~\eqref{eq:phi} and Eq.~\eqref{eq:phi_tilde} requires the calculation of the zero-energy reflection matrices $r^L$ and $r^R$.
These matrices are necessary to calculate the topological visibility $Q_\alpha = \det r^\alpha$ ($\alpha=L,R$).
For the numerical implementation of the gradient calculation, the derivative of the topological visibility is then computed using 
\begin{equation}
	\frac{ \del Q_\alpha}{\del f_{k,n}}   = Q_\alpha \tr \parS{(r^\alpha)^{-1} \frac{ \del\, r^\alpha }{\del f_{k,n}} },
\end{equation}
which follows from Jacobi's formula. This expression can be related to the derivative of a Green's function using Eq.~\eqref{eq:FisherLee} and noting that all lead quantities are independent of $f_{k,n}$. 
As this quantity depends on a single Green's function, the computational cost of the gradient of the topological visibility scales linearly with the length of the scattering region.

\subsection{Majorana performance index derivatives}
We now turn to the computation of the effective gaps defined in Eq.~\eqref{eq:h_gap}. As $\rho_j$, the zero-energy LDOS at site $j$, depends linearly on the retarded lattice Green's function $G^\ret_j$, one can use the RGF method to compute the effective gap $\tilde{\Delta}_{L(R)}$ efficiently [complexity $O(NM^3)$]. 
Since these gaps depend on $\rho_j$ on multiple sites, the computational complexity of the derivatives is {\em a priori} not obvious. Hence, we look in more detail at the calculation of the effective gap for the left-half of the system ($\tilde{\Delta}_L$). 
By symmetry, the complexity analysis will be equally valid for the right-half ($\tilde{\Delta}_R$).

The derivative of $\tilde{\Delta} _L$ with respect to some local on-site parameter $\lambda_n$ is straightforward to calculate and given by
\begin{equation}
	\frac{ \del  \tilde{\Delta}_L  }{\del \lambda_n} =  
	\frac{2  \tilde{\Delta}_L^2 }{ N \Nm_L }
	\sum_{j=1}^{N/2} \parO{ \langle x_L \rangle-j } \frac{ \del \rho_j }{\del \lambda_n}.
	\label{eq:delDelta}
\end{equation}
Using the definition of $\rho_j$ (cf. below Eq.~\eqref{eq:CM}), the preceding equation can be rewritten as a sum over derivatives of Green's function
\begin{equation}
	\frac{ \del  \tilde{\Delta}_L  }{\del \lambda_n} = 
	\frac{ \tilde{\Delta}_L^2 }{ N\pi \Nm_L }
	\mathrm{Im} \parC{  \tr \parS{  
	\sum_{j=1}^{N/2} \parO{ j - \langle x_L \rangle } \frac{ \del G^\ret_j }{\del \lambda_n}
	}}.
	\label{eq:delDelta}
\end{equation}

To lighten the notation and to make the analysis more general, we consider the efficient computation of the sum
\begin{equation}
	S_n  =  \tr \parS{ \sum_{j=1}^N \gamma_j \frac{ \del G^\ret_j }{\del \lambda_n} }, 
\end{equation}
where any other bounds on the values of $j$, such as in Eq.~\eqref{eq:delDelta}, can be implemented through the definition of $\gamma_j$. Using Eq.~\eqref{eq:dGRet}, the sum can be written as
\begin{align}
	S_n  &=  
		\tr \parS{ \sum_{j=1}^{n-1} \gamma_j G^\ret_j \frac{ \del \Sigma^R_{j+1} }{\del \lambda_n} G^\ret_j} 
		+
		\tr \parS{ \gamma_n G^\ret_n A G^\ret_n }
		\nonumber
		\\&\quad+
		\tr \parS{ \sum_{j=n+1}^N \gamma_j G^\ret_j \frac{ \del \Sigma^L_{j-1} }{\del \lambda_n} G^\ret_j},
\end{align}
where the bounds of the sums follow from the Heaviside function in Eq.~\eqref{eq:dSXP}. Considering each of these three terms separately, such that 
$S_n =  S_n^R+S_n^0 + S_n^L$,
and using the cyclic and linearity properties of the trace, one obtains
\begin{align}
	S_n^R &=   \tr \parS{M_n^R A  };
	\\
	S_n^0 &= \tr \parS{ \gamma_n \parO{G_n^\ret}^2 A};
	\\
	S_n^L &=   \tr \parS{ M_n^L A },
\end{align}
where, using Eq.~\eqref{eq:dSXP}, we have defined the matrices
\begin{align}
	M_n^R &= \sum_{j=1}^{n-1} P^R_{j+1, n}\parS{\gamma_j \parO{G_j^\ret}^2} X^R_{j+1, n},
	\\
	M_n^L &= \sum_{j=n+1}^N P^L_{j-1, n} \parS{\gamma_j \parO{G_j^\ret}^2} X^L_{j-1, n}.
\end{align}
Finally, using the definitions of the propagator strings in Eq.~\eqref{eq:rgfStrings}, one notes the recursive relations
\begin{align}
	\begin{split}
	M_{n-1}^L &= G^L_{n-1} u_{n-1}^\dag  \parS{M_{n}^L + \gamma_n \parO{G_n^\ret}^2} u_{n-1} G^L_{n-1};
		\\
		M_{n+1}^R &= G^R_{n+1} u_{n} \parS{M_{n}^R + \gamma_n \parO{G_n^\ret}^2} u_{n}^\dag G^R_{n+1},
\end{split}
	\label{eq:recursiveMLR}
\end{align}
with boundary conditions $M_{N}^L = 0$, and $M_1^R=0 $.
Since each recursion step requires a constant small number of matrix operations, calculating all the matrices $M_n^L$ and $M_n^R$, and thus the sum $S_n$ for all $n$, will have a computational complexity of $O(NM^3)$. 
This is again a polynomial speedup over the $O(N^2M^3)$ complexity that would be expected from a direct calculation of Eq.~\eqref{eq:delDelta} independently for each value of the index $n$.
Since the essential element allowing this polynomial speedup is the cyclic property of the trace, this result is valid for any sum over the LDOS at different sites. 
Going back to the analogy with the GRAPE algorithm, our result for a space integral has the same structure as the efficient calculation of a performance index that includes a time integral~\cite{boutin:2016nr}.

\begin{figure}[b]
	\centering
	\includegraphics[scale=1]{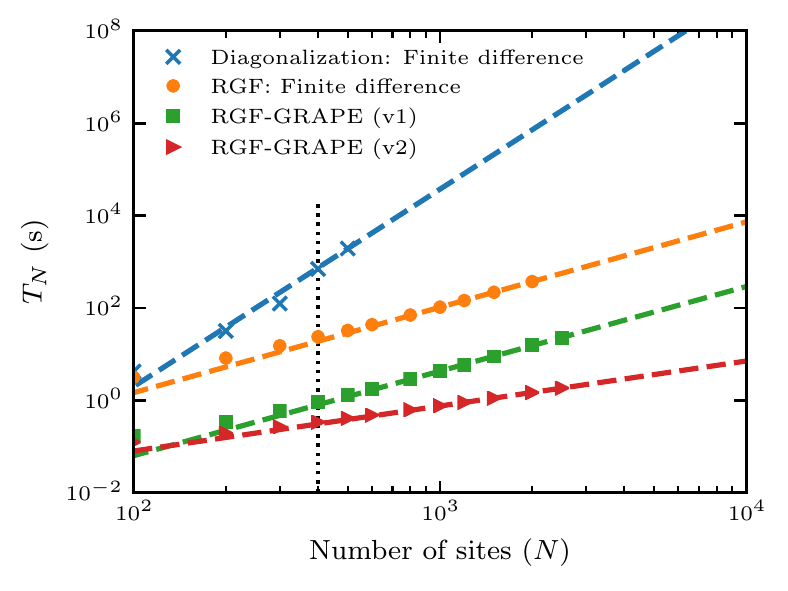}
	\caption{Average duration $T_N$ (in seconds) for the calculation of the performance index and its gradient, as a function of the length of the Majorana wire.
	The performance index is calculated using either the topological minigap (blue crosses) or the effective $p$-wave gap.
	The latter is evaluated in three different ways:
	(i) using the finite difference gradient (orange disks);
	(ii) using Eq.~\eqref{eq:delDelta} independently for each site index (green squares);
	(iii) using the recursive relation from Eq.~\eqref{eq:recursiveMLR} (red triangles).
	The last method is clearly the most efficient.
	Dashed lines are power-law fits to the numerical data (see text). 
	The dotted black vertical line indicates $N=400$, used in most of our calculations. The numerical parameters in the simulations are $\Delta = 0.0225 t$, $b = 4\Delta /3$, $\Bv_0=0$ (see Appendix~\ref{sec:modelAndParams} for more details on the model Hamiltonian).
	}
	\label{fig:benchmarkApp}
\end{figure}
\subsection{Performance of implementations}\label{subsec:bench}
To conclude this section, we supplement the previous algorithmic complexity analysis with a numerical comparison of different algorithms.
As a performance benchmark we define $T_N$, the average time used to compute the performance index and its gradient using a single core of a standard desktop computer.
To this end, we have optimized a fixed amplitude magnetic texture in a wire without intrinsic spin-orbit coupling, using different implementations of the performance index and its gradient. The texture amplitude along the $z$-axis is fixed such that the size of the optimization problem considered is $N$.
The average computation time has been obtained by dividing the total simulation time by the number of performance index evaluations carried out ($\sim 200$). The optimization is driven by an implementation of the  Broyden-Fletcher-Goldfarb-Shanno (BFGS) algorithm, where the performance index and its gradient are always computed together~\cite{Fletcher:2000uq,*Byrd:1995fv}.

Since different methods can be implemented more or less efficiently, depending on computational details such as the programming language, one should not focus much on the absolute values of $T_N$, but mainly on its scaling with $N$, the number of sites in the scattering region. To this end, one can consider this scaling quantitatively using a power-law fit to the numerical data such that $T_N \propto N^\xi$. The exponent $\xi$ should then be compared to the expected complexity
\footnote{
As we consider a single subband nanowire, the number of local degree of freedoms is fixed to $M=4$ in all calculations. Hence, contrary to previous sections, we do not consider the value of $M$ in the complexity considerations. All methods considered should have a complexity $O(N^\xi M^3)$.}.

Figure~\ref{fig:benchmarkApp} compares $T_N$ for 4 algorithms and performance indices as a function of $N$. For each method, the dashed curve is the result of a power-law fit. 
The blue crosses are computing times for the optimization of the Majorana performance index $\Phi' = -\delta_E Q_L$ (used in Appendix~\ref{sec:obcSweep}), where the minigap $\delta_E$ is obtained through diagonalization of the isolated superconducting region Hamiltonian and the gradient is computed using finite difference
\footnote{The minigap is defined as the lowest excitation energy of the Majorana wire decoupled from the leads. In the main text, we refer to it as the \emph{topological gap} or simply the \emph{energy gap}. }. 
The three other datasets are computing times for the effective performance index, Eq.~\eqref{eq:phi_tilde}, where we use the RGF method to compute the performance index. Three different methods were used to compute the gradient of the performance index : (i) finite difference (orange disks), (ii) Eq.~\eqref{eq:delDelta} independently for each site index (green squares, labeled RGF-GRAPE v1), and (iii) the recursive relations of Eq.~\eqref{eq:recursiveMLR} (red triangles, labeled RGF-GRAPE v2). These three gradient calculation methods lead to the same gradient up to numerical precision. 

We now turn to the fit results. In the case of the diagonalization, we extract the exponent $\xi \approx 4.27$, which is close to the expected complexity $O(N^4)$
\footnote{Matrix diagonalization is $O(N^3)$ and finite difference requires to perform $N$ of them for the problem considered.}. 
For the RGF-based calculations, all obtained exponents are below 2 (respectively $\xi \approx 1.85, \, 1.83,$ and 0.98), showing that, independent of the details of the implementation of the gradient there is a clear advantage from the computational point-of-view to consider the effective gap instead of the minigap. In addition, these exponents are in agreement with the complexity analysis of Sec.~\ref{subsec:RGFopt}, which stated that the calculation of the gradient of a sum over $N$ sites should be at worst $O(N^2)$. Finally, we note that the use of the recursion relations of Eq.~\eqref{eq:recursiveMLR} allows to reach a linear complexity, which is a polynomial speedup over finite difference. 
 
Finally, to put these results into perspective, one can also look at the actual computation times for the different methods. As an example, for $N=400$ (dotted black vertical line) the calculation times of the performance index and its gradient are $T_N = 0.34,\, 0.91,\, 23,$ and~700~s. Although one should use caution when interpreting such results, since absolute timings are implementation-dependent, these numbers help to put in perspective the concrete advantage of reducing the complexity of the gradient calculation.
Indeed, in the context of an optimization algorithm, which requires to repeat this calculation hundreds if not thousands of times, a speedup of the gradient calculation allows, for a fixed computation time, to consider either more realistic wire models, or more sophisticated optimization algorithms that get closer to a global maximum.

\section{Nanowire model and parameters}\label{sec:modelAndParams}
Following the notation of Eq.~\eqref{eq:hTB}, we consider in our calculations a noninteracting tight-binding model
for a unidimensional wire (single subband) with a superconducting scattering region of $N$ sites ($j = 1, \dots N$) coupled to semi-infinite metallic leads (sites $j=0,-1, \dots$ for the left lead and sites $j=N+1, N+2, \dots$ for the right lead). 
In this case, the number of local degrees of freedom is $M=4$, and the spinors are taken in a Bogoliubov-de Gennes basis such that
\begin{equation}
	\psi^\dag_j = \parO{c^\dag_{j,\upa}, c^\dag_{j,\da} , - c_{j,\da}, c_{j,\upa}  },
\end{equation}
with $c^{(\dag)}_{j,\sigma}$, an operator annihilating (creating) a fermion at site $j$  with spin $\sigma$.
Denoting $\tau_\alpha$ ($\sigma_\alpha$) the Pauli matrices acting on the particle-hole (spin) sectors, the structure of 
the hopping matrices is
\begin{equation}
	u_j = -t \tz  - i  \alpha \sy \tz,
\end{equation}
where $t=\hbar^2/2m^* a^2$ is the hopping amplitude with $m^*$ the effective mass and $a$ the lattice constant, and $\alpha$ is the spin-orbit coupling amplitude.
The onsite Hamiltonian reads
\begin{equation}
	h_j = (2t-\mu_j)\tz  + (\Bv_0 + \dB_j) \cdot \sigv + \Delta_j \tx  
	,
\end{equation}
where $\mu_j$ is the effective local potential including both the local electrostatic gate voltage and the chemical potential, 
$\Delta_j$ is the local proximity-induced superconducting gap, 
$\Bv_0$ is a uniform magnetic field, and $\dB_j$ is a possibly non-uniform local magnetic field. 
All prefactors relating the magnetic field to the Zeeman energy, including the $g$-factor,  are absorbed in the definitions of $\Bv_0$ and $\dB_j$.

Unless otherwise stated, in all numerics,
we consider approximate parameters for a semiconductor with negligible intrinsic spin-orbit coupling ($\alpha=0$)~\footnote{
	The exception is Fig.~\ref{mu_modulation}, where we take the parameters $\alpha = 0.05 t$, $\Delta=0.0225 t$ and $\Bv_0/t = 0.027 \hat \xv$.
}.
We take a lattice constant $a=10$~nm ($N=200$ then corresponds to a 2~$\mu$m long nanowire) and an effective mass $m^* = 0.2\,m_e$, where $m_e$ is the bare electron mass. Those parameters lead to a hopping amplitude $t=1.9$~meV.
Based on the experimentally observed proximity-induced superconductivity in a GaAs two-dimensional electron gas~\cite{Wan:2015qv}, we take $\Delta= 0.0225t$ ($43\,\mu$eV). For a $g$-factor of 2, $|\Bv_0 + \dB| = 4 \Delta /3 \approx 0.03t$ corresponds to a magnetic field of 1~Tesla.
In addition, we consider uniform metallic leads ($\Delta=0$), with a large density of states ($\mu= 1.9t$) and strongly coupled to the superconducting region (no barrier at the interface).
The same values of $t$, $\alpha$ and $\Bv_0$ are used in both the leads and the superconducting scattering region.
The strong coupling to the leads was found to ensure a good convergence of the optimization algorithm. However, as illustrated by the results of diagonalization shown in the main text (see e.g. Fig~\ref{fig:phaseSpaceExample}(d)), which correspond to the regime of isolated wires, the final results appear to be robust with respect to the details of the lead couplings and parameters.
Finally, to improve the numerical stability of recursive calculations, the zero-energy Green's functions are calculated at a small but finite imaginary energy $E/t=10^{-6} \,i $.

\begin{figure}[tb]
	\centering
	\includegraphics[width=\linewidth]{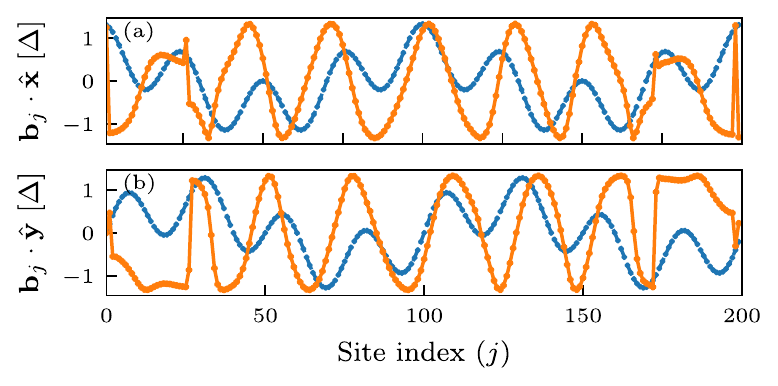}
	\caption{(a) the $x-$ and (b) the $y$-components of the initial (dashed blue) and the optimized (solid orange) magnetic textures without penalties for smoothness. 
	}
	\label{fig:optExample}
\end{figure}

\section{Regularization and parameter scaling}\label{regularization}
In order to favor smooth spatial profiles, penalty functions can be added to the performance index. 
We refer the reader to Ref.~\cite{Leung:2017lh} and references therein for example penalties used in time-domain optimizations. 
In the case of a scalar quantity such as the electrostatic gate voltage, we add a penalty function
\begin{equation}
	\Phi_\mu = - \beta_\mu  \sum_{j=1}^{N-1} \parO{ \mu_j - \mu_{j+1} }^2,
\end{equation}
where $\beta_\mu>0$ is a parameter that weights the cost of spatial variations in the gate voltage.
A large weight will favor a flat voltage profile independently of the problem.

This type of penalty can be generalized to a vector profile. In the case of the fixed amplitude magnetic texture, we use the penalty
\begin{equation}
	\Phi_{b} =  \beta_{b} \sum_{j=1}^{N-1} \parS{\hat{\textbf{b}}_j\cdot \hat{\textbf{b}}_{j+1} -1},
\end{equation}
where $\hat{\dB}_j = \dB_j/|\dB_j|$ is a unit vector and $\beta_{b}>0$ is again a weight factor for the penalty. This penalty will favor a smooth and uniform magnetic texture, which should be more easily realizable experimentally.

In the case of an optimization involving parameters with different units or scales, it can be useful to introduce a scaling parameter $w$ in order to control the relative weight of the different types of parameters in the gradient descent. In particular, in the case of Fig.~\ref{fig:phaseSpace} where the orientation of a fixed amplitude magnetic texture and a uniform chemical potential are optimized, a parameter $w_\mu$ is introduced in order to reduce the relative weight of the chemical potential in the optimization. This was found to reduce the risk of the optimizer (BFGS) declaring convergence too quickly following a large reduction of the gradient of the performance index with respect to only the chemical potential, without much change in the magnetic texture.

The optimal value of the heuristic parameters $\beta$ and $w$ depends on the details of the problem. Since these are not known, it is useful to perform the optimization for a few different values.
In particular, the results of Fig.~\ref{fig:phaseSpace} are obtained by considering the solution with the largest minigap $\delta_E$ out of 8 optimizations. These 8 runs of the algorithm, all starting from the same initial spiral magnetic texture, differed by the choice of the optimization parameters $\beta_b$ and $w_\mu$ which were taken to be the possible combinations of $w_\mu \in \parC{10^{-2}, 10^{-4}}$ and $\beta_b \in \parC{ 0.01, 0.1, 0.5, 1}$. Due to the variation of the problem landscape and the performance index amplitude, different parameters performed better in some regions of the ($B_0$, $b$) parameter space than others. These values were found sufficient to obtain improved MBS characteristics compared to the initial configuation in all region of parameter space.

\begin{figure}[tb]
	\centering
	\includegraphics[width=\linewidth]{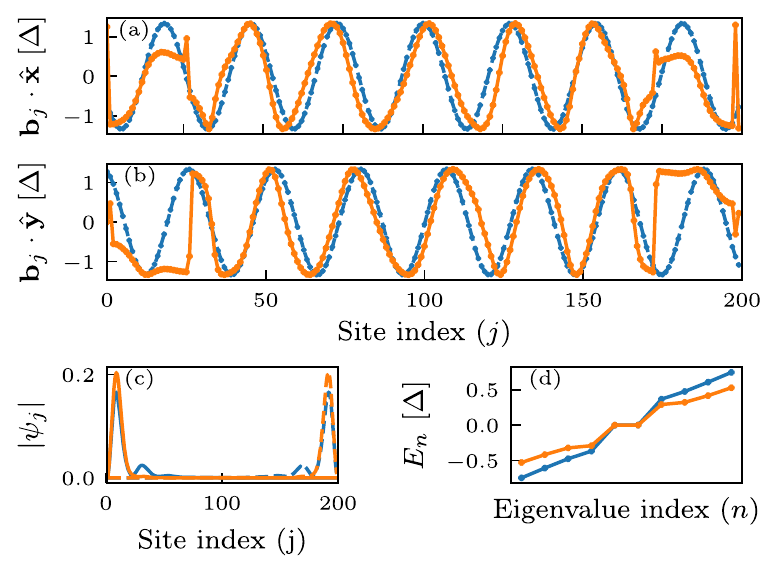}
	\caption{(a) the $x-$ and (b) the $y$-components for the 
	optimized (solid orange) magnetic textures. Dashed blue curves are a spiral texture obtained by a harmonic fit to the $x$-component of the optimized texture.
	(c) Wavefunction amplitude for the left and right MBS (solid and dashed curves, respectively).
	(d) Spectrum of the isolated superconducting scattering region close to zero energy.
	For all panels, the orange curves are results for the optimized texture, while the blue curves are results for a perfect spiral texture fitted to the optimized texture.
	}
	\label{fig:boundary}
\end{figure}
%%%%%%%%%%%%%%%%%%%%%%%%%%%%%%%%%%%%%%%%%%%%%%%%%%%%%%%%%%%%%%%%%%%%%%%%%%%%%%%
% Boundary effects and reality checks
%%%%%%%%%%%%%%%%%%%%%%%%%%%%%%%%%%%%%%%%%%%%%%%%%%%%%%%%%%%%%%%%%%%%%%%%%%%%%%%
% 
\section{Proof of principle for the optimization algorithm and the role of boundaries}\label{boundaries}

In this appendix, we expand on the optimization results presented as a proof of concept in Fig.~\ref{fig1}c.
For this optimization, we consider a superconducting nanowire with $N=200$ sites, with neither intrinsic spin-orbit interaction ($\alpha = 0$) nor external magnetic field ($\Bv_0 = 0)$. We optimize both the local amplitude and orientation of a magnetic texture (with the constraints $b_j \leq 0.03t = 4 \Delta/3$ and $\dB_j\cdot \hat \zv = 0$). No penalties for smoothness are added ($\beta_b =0$). In addition, we optimize the chemical potential in the superconducting region (though we restrict ourselves to a spatially uniform chemical potential).

Figure~\ref{fig:optExample} compares the initial magnetic texture to the optimized result. Starting from an initial texture consisting of a sum of two spirals of different periods, the optimization converges to a solution which is spiral-like in the bulk (panels a and b), with a uniform amplitude $b_j = 4\Delta/3$ (the maximal value allowed by the imposed constraints). As shown in Fig.~\ref{fig1}c, the system is initially in the trivial phase and the optimization drives the parameters through a topological phase transition.
It is worth noting that the optimization naturally finds a smooth solution in the bulk even though no penalties were used in this optimization. By introducing such a penalty, discontinuities near the boundary can be reduced (not shown).

In order to better understand the role of the boundaries of the optimized texture, we fit the $x$-component of the optimized magnetic texture to an harmonic function $\dB_j \cdot \hat \xv = A \cos( 2\pi j/R + \phi)$ (with fit parameters $A$, $R$ and $\phi$). Fig.~\ref{fig:boundary} compares the optimized texture to a spiral texture built from the fit result. One can see that, while the spiral texture has a larger minigap (0.37~$\Delta$ compared to 0.29~$\Delta$), the MBS in the case of the optimized texture is more localized. This smaller localization length leads to a reduced overlap of the Majorana wavefunctions and thus to a reduced zero-mode splitting from $\epsilon_M = 1 \times 10^{-3}$ to $\epsilon_M  = 2 \times 10^{-5}$.

\begin{figure}[tb]
	\centering
	\includegraphics[width=\linewidth]{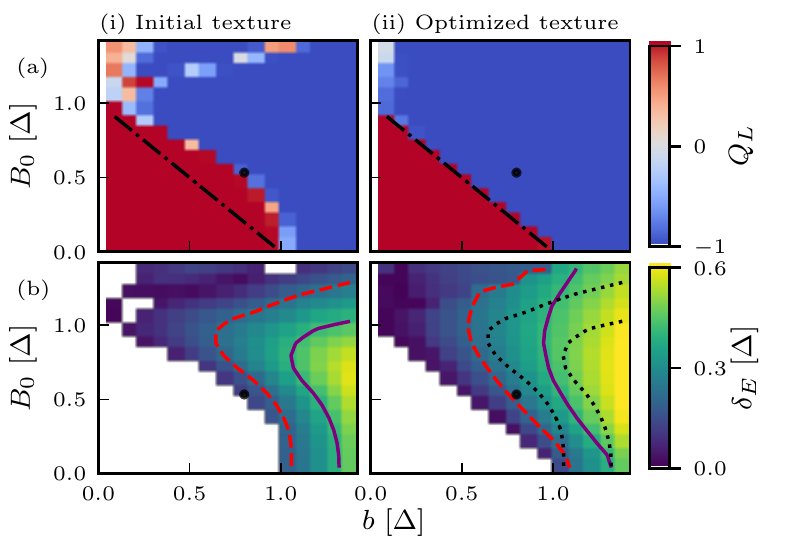}
	\caption{(a,b) Topological visibility in a nanowire without spin-orbit coupling. (a) Initial spiral magnetic texture. (b) Magnetic textures optimized using Eq.~\eqref{phiMinigap}. 
	(c,d) Energy minigap $\delta_E$. (c) Initial spiral magnetic texture. (d) Optimized magnetic texture.
	Purple dashed (solid green) curves denote minigaps $\delta_E = 100$(200)~mK.
	Dotted black contours are a copy of the contours for the initial spiral texture (panel to the left).
	The dot-dashed black line indicates $B_0 + b = \Delta$, the expected
	minimal Zeeman energy to reach the topological phase.
	}
	\label{fig:exact_OBC}
\end{figure}
\begin{figure}[tb]
	\centering
	\includegraphics[width=\linewidth]{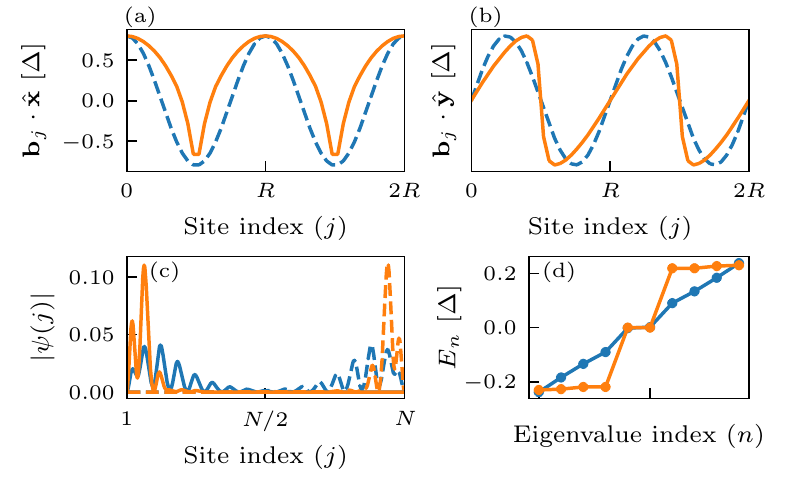}
	\caption{Non-optimized (spiral, dashed blue) and optimized (solid orange) curves for (a) the $x-$ and (b) the $y$-components of the magnetic texture. Optimization was performed for a fixed amplitude texture $|\dB_j|= 0.018t = 4\Delta/5 $ and a uniform magnetic field $\Bv_0/t =0.012\, \hat{\mathbf{x}}$ (black dot in Fig.~\ref{fig:exact_OBC}).
	(c) Solid (dashed) curves indicate the left (right) MBS' wavefunctions. 
	For all panels, blue curves are the results for the initial spiral texture, while the orange curves are the results for the optimized texture.
	(d) Spectrum of the isolated superconducting scattering region close to zero energy.
	}
	\label{fig:fields}
\end{figure}

\section{Maximizing the topological minigap}	\label{sec:obcSweep}
As a complement to the results of Sec.~\ref{ssec:texture}, we consider the same magnetic texture optimization, but using the performance index   
 \begin{equation}
 	\Phi' = -\delta_E Q_L,
 	\label{phiMinigap}
 \end{equation}
 where $\delta_E$ is the minigap obtained by diagonalization of the single-particle Hamiltonian~\footnote{We consider again a scattering region of length $N=400$, and constrain the optimization problem to textures with a  periodicity of $25$ sites and $\dB_j \cdot \hat{\mathbf{z}}=0$.}.

Figure~\ref{fig:exact_OBC} presents optimization results for a grid of points in the $(B_0, b$) parameter space. Panel (a) compares
the topological visibility for an initial spiral texture (column (i)) to an optimized texture (column (ii)). 
As for Fig.~\ref{fig:phaseSpace}, where the more numerically efficient performance index $\tilde{\Phi}$ is optimized, the area of the topological (blue) region of parameter space increases for the optimized texture. For both performance indices, the optimization leads to the topological phase almost independently of the position in parameter space, as long as the total Zeeman energy $B_0 + b$ is larger than the superconducting gap $\Delta$. 
Panel (b) presents the minigap $\delta_E$ in the topological phase. 
Similarly to the optimization of $\tilde{\Phi}$ in Fig~\ref{fig:phaseSpace}, the direct optimization of the minigap leads to constant gap contours enclosing a larger area of parameter space for the topological phase.

Finally, we compare the optimization results for a point in parameter space (corresponding to the black disks in Fig.~\ref{fig:exact_OBC}). 
Figure~\ref{fig:fields}(a,b) shows the components of the optimized magnetic texture for two periods (solid orange curve) and compares them to the initial spiral texture (dashed blue curve). Figure~\ref{fig:fields}(c,d) displays both the MBS' wavefunction (panel c) and the spectrum for the superconducting region decoupled from the leads (panel d). The optimization leads to both an increased minigap and enhanced energy degeneracy between the left and right MBS. These optimization results are equivalent to the results obtained by the optimization of $\tilde{\Phi}$ presented in Fig.~\ref{fig:phaseSpaceExample}. 
Hence, similar results to the numerically costly exact optimization of the topological minigap can be obtained using the LDOS-based effective performance index~$\tilde{\Phi}$ presented in Sec.~\ref{sec:majOpt}.

\begin{figure}[tb]
	\centering
	\includegraphics[width=\linewidth]{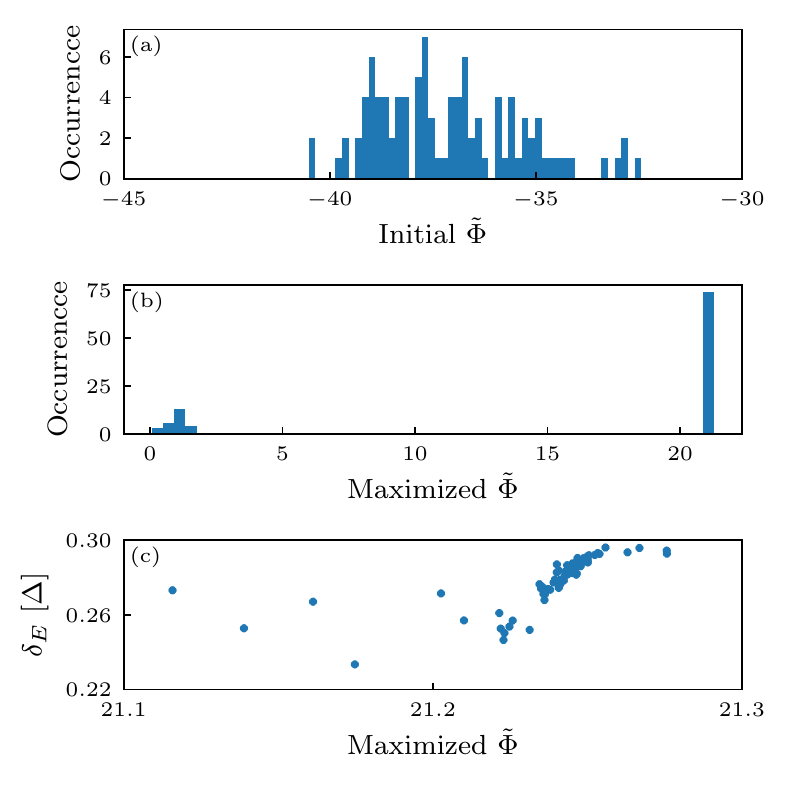}
	\caption{Statistics of convergence of the RGF-GRAPE algorithm for the optimization of a fixed amplitude magnetic texture starting from 100 different random texture realizations (see text).
	(a) Distribution of the initial value of the performance index.
	(b) Distribution of the final value of the performance index after running the BFGS optimization algorithm until convergence.
	(c) Scatter plot of the bulk gap (obtained through diagonalization) 
	 as a function of the final (maximized) value of the performance index.
	A uniform chemical potential in the wire is also optimized (initial value $\mu=10^{-3} t$, $w_\mu=10^{-3}$). The numerical parameters in the simulations are $N=200$, $b = 4\Delta/3$, $\Delta = 0.0225 t$, $B_0 = \alpha=\beta_b=0$.
	}
	\label{fig:histogram}
\end{figure}

\section{Convergence statistics}\label{convergence}
In this appendix, we present additional data related to the convergence of the RGF-GRAPE algorithm and the bulk gap of the optimization results. 
To this end, we performed 100 optimizations each starting from a different realization of a fixed-amplitude random magnetic texture.
In each realization, the magnetic field at site $j$ is 
	$\mathbf{b}_j(\theta_j, \phi_j) = b(\sin \theta_j \cos \phi_j \hat{\mathbf{x}} + \sin \theta_j \cos \phi_j \hat{\mathbf{y}} + \cos \theta_j \hat{\mathbf{z}})$,
where $\phi_j$ and $\theta_j$ are uncorrelated random variables taken from a uniform distribution of width $2\pi$ respectively centered around $0$ and $\pi/2$. 

Figure~\ref{fig:histogram}(a) presents the distribution of the initial value of the performance index. The negative values indicate that all initial textures are in the trivial phase. Panel (b) shows the distribution of the final performance index after performing an optimization using the BFGS algorithm where the gradient of $\tilde{\Phi}$ was calculated at each iteration using RGF-GRAPE. 
For 26 occurrences the optimization converged to a solution near the topological phase transition ($|\tilde{\Phi}|<2$), 
where the performance index value is dominated by the topological visibility. 
In the remaining 74 cases the optimization converged deep in the topological phase. 
For these occurrences, panel (c) shows a scatter plot of the resulting bulk gap
where we can see a relation between the value of the performance index and the bulk gap. 
For the 26 cases with $|\tilde{\Phi}|<2$, the failure to find solutions deep in the topological phase might be explained by the presence of local minima in the high-dimensional optimization space. This standard problem of optimization can we be solved by using standard methods of global optimization such as the basin hopping method.

\bibliographystyle{apsrev4-1}
% \bibliography{bibliography}

\begin{thebibliography}{81}%
\makeatletter
\providecommand \@ifxundefined [1]{%
 \@ifx{#1\undefined}
}%
\providecommand \@ifnum [1]{%
 \ifnum #1\expandafter \@firstoftwo
 \else \expandafter \@secondoftwo
 \fi
}%
\providecommand \@ifx [1]{%
 \ifx #1\expandafter \@firstoftwo
 \else \expandafter \@secondoftwo
 \fi
}%
\providecommand \natexlab [1]{#1}%
\providecommand \enquote  [1]{``#1''}%
\providecommand \bibnamefont  [1]{#1}%
\providecommand \bibfnamefont [1]{#1}%
\providecommand \citenamefont [1]{#1}%
\providecommand \href@noop [0]{\@secondoftwo}%
\providecommand \href [0]{\begingroup \@sanitize@url \@href}%
\providecommand \@href[1]{\@@startlink{#1}\@@href}%
\providecommand \@@href[1]{\endgroup#1\@@endlink}%
\providecommand \@sanitize@url [0]{\catcode `\\12\catcode `\$12\catcode
  `\&12\catcode `\#12\catcode `\^12\catcode `\_12\catcode `\%12\relax}%
\providecommand \@@startlink[1]{}%
\providecommand \@@endlink[0]{}%
\providecommand \url  [0]{\begingroup\@sanitize@url \@url }%
\providecommand \@url [1]{\endgroup\@href {#1}{\urlprefix }}%
\providecommand \urlprefix  [0]{URL }%
\providecommand \Eprint [0]{\href }%
\providecommand \doibase [0]{http://dx.doi.org/}%
\providecommand \selectlanguage [0]{\@gobble}%
\providecommand \bibinfo  [0]{\@secondoftwo}%
\providecommand \bibfield  [0]{\@secondoftwo}%
\providecommand \translation [1]{[#1]}%
\providecommand \BibitemOpen [0]{}%
\providecommand \bibitemStop [0]{}%
\providecommand \bibitemNoStop [0]{.\EOS\space}%
\providecommand \EOS [0]{\spacefactor3000\relax}%
\providecommand \BibitemShut  [1]{\csname bibitem#1\endcsname}%
\let\auto@bib@innerbib\@empty
%</preamble>
\bibitem [{\citenamefont {Mourik}\ \emph {et~al.}(2012)\citenamefont {Mourik},
  \citenamefont {Zuo}, \citenamefont {Frolov}, \citenamefont {Plissard},
  \citenamefont {Bakkers},\ and\ \citenamefont {Kouwenhoven}}]{Mourik:2012ee}%
  \BibitemOpen
  \bibfield  {author} {\bibinfo {author} {\bibfnamefont {V.}~\bibnamefont
  {Mourik}}, \bibinfo {author} {\bibfnamefont {K.}~\bibnamefont {Zuo}},
  \bibinfo {author} {\bibfnamefont {S.~M.}\ \bibnamefont {Frolov}}, \bibinfo
  {author} {\bibfnamefont {S.}~\bibnamefont {Plissard}}, \bibinfo {author}
  {\bibfnamefont {E.}~\bibnamefont {Bakkers}}, \ and\ \bibinfo {author}
  {\bibfnamefont {L.}~\bibnamefont {Kouwenhoven}},\ }\href@noop {} {\bibfield
  {journal} {\bibinfo  {journal} {Science}\ }\textbf {\bibinfo {volume}
  {336}},\ \bibinfo {pages} {1003} (\bibinfo {year} {2012})}\BibitemShut
  {NoStop}%
\bibitem [{\citenamefont {Deng}\ \emph {et~al.}(2016)\citenamefont {Deng},
  \citenamefont {Vaitiekenas}, \citenamefont {Hansen}, \citenamefont {Danon},
  \citenamefont {Leijnse}, \citenamefont {Flensberg}, \citenamefont
  {Nyg{\aa}rd}, \citenamefont {Krogstrup},\ and\ \citenamefont
  {Marcus}}]{Deng:2016wt}%
  \BibitemOpen
  \bibfield  {author} {\bibinfo {author} {\bibfnamefont {M.~T.}\ \bibnamefont
  {Deng}}, \bibinfo {author} {\bibfnamefont {S.}~\bibnamefont {Vaitiekenas}},
  \bibinfo {author} {\bibfnamefont {E.~B.}\ \bibnamefont {Hansen}}, \bibinfo
  {author} {\bibfnamefont {J.}~\bibnamefont {Danon}}, \bibinfo {author}
  {\bibfnamefont {M.}~\bibnamefont {Leijnse}}, \bibinfo {author} {\bibfnamefont
  {K.}~\bibnamefont {Flensberg}}, \bibinfo {author} {\bibfnamefont
  {J.}~\bibnamefont {Nyg{\aa}rd}}, \bibinfo {author} {\bibfnamefont
  {P.}~\bibnamefont {Krogstrup}}, \ and\ \bibinfo {author} {\bibfnamefont
  {C.~M.}\ \bibnamefont {Marcus}},\ }\href {\doibase 10.1126/science.aaf3961}
  {\bibfield  {journal} {\bibinfo  {journal} {Science}\ }\textbf {\bibinfo
  {volume} {354}},\ \bibinfo {pages} {1557} (\bibinfo {year}
  {2016})}\BibitemShut {NoStop}%
\bibitem [{\citenamefont {Chen}\ \emph {et~al.}(2017)\citenamefont {Chen},
  \citenamefont {Yu}, \citenamefont {Stenger}, \citenamefont {Hocevar},
  \citenamefont {Car}, \citenamefont {Plissard}, \citenamefont {Bakkers},
  \citenamefont {Stanescu},\ and\ \citenamefont {Frolov}}]{Chene2017}%
  \BibitemOpen
  \bibfield  {author} {\bibinfo {author} {\bibfnamefont {J.}~\bibnamefont
  {Chen}}, \bibinfo {author} {\bibfnamefont {P.}~\bibnamefont {Yu}}, \bibinfo
  {author} {\bibfnamefont {J.}~\bibnamefont {Stenger}}, \bibinfo {author}
  {\bibfnamefont {M.}~\bibnamefont {Hocevar}}, \bibinfo {author} {\bibfnamefont
  {D.}~\bibnamefont {Car}}, \bibinfo {author} {\bibfnamefont {S.~R.}\
  \bibnamefont {Plissard}}, \bibinfo {author} {\bibfnamefont {E.~P. A.~M.}\
  \bibnamefont {Bakkers}}, \bibinfo {author} {\bibfnamefont {T.~D.}\
  \bibnamefont {Stanescu}}, \ and\ \bibinfo {author} {\bibfnamefont {S.~M.}\
  \bibnamefont {Frolov}},\ }\href
  {http://advances.sciencemag.org/content/3/9/e1701476} {\bibfield  {journal}
  {\bibinfo  {journal} {Science Advances}\ }\textbf {\bibinfo {volume} {3}},\
  \bibinfo {pages} {e1701476} (\bibinfo {year} {2017})}\BibitemShut {NoStop}%
\bibitem [{\citenamefont {Zhang}\ \emph {et~al.}(2018)\citenamefont {Zhang},
  \citenamefont {Liu}, \citenamefont {Gazibegovic}, \citenamefont {Xu},
  \citenamefont {Logan}, \citenamefont {Wang}, \citenamefont {van Loo},
  \citenamefont {Bommer}, \citenamefont {de~Moor}, \citenamefont {Car},
  \citenamefont {Op~het Veld}, \citenamefont {van Veldhoven}, \citenamefont
  {Koelling}, \citenamefont {Verheijen}, \citenamefont {Pendharkar},
  \citenamefont {Pennachio}, \citenamefont {Shojaei}, \citenamefont {Lee},
  \citenamefont {Palmstrøm}, \citenamefont {Bakkers}, \citenamefont {Sarma},\
  and\ \citenamefont {Kouwenhoven}}]{Zhang2018a}%
  \BibitemOpen
  \bibfield  {author} {\bibinfo {author} {\bibfnamefont {H.}~\bibnamefont
  {Zhang}}, \bibinfo {author} {\bibfnamefont {C.-X.}\ \bibnamefont {Liu}},
  \bibinfo {author} {\bibfnamefont {S.}~\bibnamefont {Gazibegovic}}, \bibinfo
  {author} {\bibfnamefont {D.}~\bibnamefont {Xu}}, \bibinfo {author}
  {\bibfnamefont {J.~A.}\ \bibnamefont {Logan}}, \bibinfo {author}
  {\bibfnamefont {G.}~\bibnamefont {Wang}}, \bibinfo {author} {\bibfnamefont
  {N.}~\bibnamefont {van Loo}}, \bibinfo {author} {\bibfnamefont {J.~D.~S.}\
  \bibnamefont {Bommer}}, \bibinfo {author} {\bibfnamefont {M.~W.~A.}\
  \bibnamefont {de~Moor}}, \bibinfo {author} {\bibfnamefont {D.}~\bibnamefont
  {Car}}, \bibinfo {author} {\bibfnamefont {R.~L.~M.}\ \bibnamefont {Op~het
  Veld}}, \bibinfo {author} {\bibfnamefont {P.~J.}\ \bibnamefont {van
  Veldhoven}}, \bibinfo {author} {\bibfnamefont {S.}~\bibnamefont {Koelling}},
  \bibinfo {author} {\bibfnamefont {M.~A.}\ \bibnamefont {Verheijen}}, \bibinfo
  {author} {\bibfnamefont {M.}~\bibnamefont {Pendharkar}}, \bibinfo {author}
  {\bibfnamefont {D.~J.}\ \bibnamefont {Pennachio}}, \bibinfo {author}
  {\bibfnamefont {B.}~\bibnamefont {Shojaei}}, \bibinfo {author} {\bibfnamefont
  {J.~S.}\ \bibnamefont {Lee}}, \bibinfo {author} {\bibfnamefont {C.~J.}\
  \bibnamefont {Palmstrøm}}, \bibinfo {author} {\bibfnamefont {E.~P. A.~M.}\
  \bibnamefont {Bakkers}}, \bibinfo {author} {\bibfnamefont {S.~D.}\
  \bibnamefont {Sarma}}, \ and\ \bibinfo {author} {\bibfnamefont {L.~P.}\
  \bibnamefont {Kouwenhoven}},\ }\href {http://dx.doi.org/10.1038/nature26142}
  {\bibfield  {journal} {\bibinfo  {journal} {Nature}\ }\textbf {\bibinfo
  {volume} {556}},\ \bibinfo {pages} {74} (\bibinfo {year} {2018})}\BibitemShut
  {NoStop}%
\bibitem [{\citenamefont {Kitaev}(2001)}]{Kitaev:2001sp}%
  \BibitemOpen
  \bibfield  {author} {\bibinfo {author} {\bibfnamefont {A.~Y.}\ \bibnamefont
  {Kitaev}},\ }\href {http://stacks.iop.org/1063-7869/44/i=10S/a=S29}
  {\bibfield  {journal} {\bibinfo  {journal} {Physics-Uspekhi}\ }\textbf
  {\bibinfo {volume} {44}},\ \bibinfo {pages} {131} (\bibinfo {year}
  {2001})}\BibitemShut {NoStop}%
\bibitem [{\citenamefont {Oreg}\ \emph {et~al.}(2010)\citenamefont {Oreg},
  \citenamefont {Refael},\ and\ \citenamefont {{{von Oppen}}}}]{Oreg:2010xy}%
  \BibitemOpen
  \bibfield  {author} {\bibinfo {author} {\bibfnamefont {Y.}~\bibnamefont
  {Oreg}}, \bibinfo {author} {\bibfnamefont {G.}~\bibnamefont {Refael}}, \ and\
  \bibinfo {author} {\bibfnamefont {F.}~\bibnamefont {{{von Oppen}}},
  \bibfnamefont {Felix}},\ }\href {\doibase 10.1103/PhysRevLett.105.177002}
  {\bibfield  {journal} {\bibinfo  {journal} {Phys. Rev. Lett.}\ }\textbf
  {\bibinfo {volume} {105}},\ \bibinfo {pages} {177002} (\bibinfo {year}
  {2010})}\BibitemShut {NoStop}%
\bibitem [{\citenamefont {Pientka}\ \emph {et~al.}(2015)\citenamefont
  {Pientka}, \citenamefont {Peng}, \citenamefont {Glazman},\ and\ \citenamefont
  {von Oppen}}]{Pientka2015review}%
  \BibitemOpen
  \bibfield  {author} {\bibinfo {author} {\bibfnamefont {F.}~\bibnamefont
  {Pientka}}, \bibinfo {author} {\bibfnamefont {Y.}~\bibnamefont {Peng}},
  \bibinfo {author} {\bibfnamefont {L.}~\bibnamefont {Glazman}}, \ and\
  \bibinfo {author} {\bibfnamefont {F.}~\bibnamefont {von Oppen}},\ }\href
  {http://stacks.iop.org/1402-4896/2015/i=T164/a=014008} {\bibfield  {journal}
  {\bibinfo  {journal} {Physica Scripta}\ }\textbf {\bibinfo {volume} {2015}},\
  \bibinfo {pages} {014008} (\bibinfo {year} {2015})}\BibitemShut {NoStop}%
\bibitem [{\citenamefont {Alicea}(2012)}]{alicea:2012fj}%
  \BibitemOpen
  \bibfield  {author} {\bibinfo {author} {\bibfnamefont {J.}~\bibnamefont
  {Alicea}},\ }\href@noop {} {\bibfield  {journal} {\bibinfo  {journal}
  {Reports on Progress in Physics}\ }\textbf {\bibinfo {volume} {75}},\
  \bibinfo {pages} {076501} (\bibinfo {year} {2012})}\BibitemShut {NoStop}%
\bibitem [{\citenamefont {Beenakker}(2013)}]{Beenakker:2013ve}%
  \BibitemOpen
  \bibfield  {author} {\bibinfo {author} {\bibfnamefont {C.}~\bibnamefont
  {Beenakker}},\ }\href {\doibase 10.1146/annurev-conmatphys-030212-184337}
  {\bibfield  {journal} {\bibinfo  {journal} {Annual Review of Condensed Matter
  Physics}\ }\textbf {\bibinfo {volume} {4}},\ \bibinfo {pages} {113} (\bibinfo
  {year} {2013})}\BibitemShut {NoStop}%
\bibitem [{\citenamefont {Aguado}(2017)}]{AguadoReview}%
  \BibitemOpen
  \bibfield  {author} {\bibinfo {author} {\bibfnamefont {R.}~\bibnamefont
  {Aguado}},\ }\href {\doibase 10.1393/ncr/i2017-10141-9} {\bibfield  {journal}
  {\bibinfo  {journal} {La Rivista del Nuovo Cimento}\ }\textbf {\bibinfo
  {volume} {40}},\ \bibinfo {pages} {523–593} (\bibinfo {year}
  {2017})}\BibitemShut {NoStop}%
\bibitem [{\citenamefont {Lutchyn}\ \emph {et~al.}(2018)\citenamefont
  {Lutchyn}, \citenamefont {Bakkers}, \citenamefont {Kouwenhoven},
  \citenamefont {Krogstrup}, \citenamefont {Marcus},\ and\ \citenamefont
  {Oreg}}]{Lutchyn2017}%
  \BibitemOpen
  \bibfield  {author} {\bibinfo {author} {\bibfnamefont {R.}~\bibnamefont
  {Lutchyn}}, \bibinfo {author} {\bibfnamefont {E.}~\bibnamefont {Bakkers}},
  \bibinfo {author} {\bibfnamefont {L.}~\bibnamefont {Kouwenhoven}}, \bibinfo
  {author} {\bibfnamefont {P.}~\bibnamefont {Krogstrup}}, \bibinfo {author}
  {\bibfnamefont {C.}~\bibnamefont {Marcus}}, \ and\ \bibinfo {author}
  {\bibfnamefont {Y.}~\bibnamefont {Oreg}},\ }\href@noop {} {\bibfield
  {journal} {\bibinfo  {journal} {Nature Reviews Materials}\ }\textbf {\bibinfo
  {volume} {3}},\ \bibinfo {pages} {52} (\bibinfo {year} {2018})}\BibitemShut
  {NoStop}%
\bibitem [{\citenamefont {Choy}\ \emph {et~al.}(2011)\citenamefont {Choy},
  \citenamefont {Edge}, \citenamefont {Akhmerov},\ and\ \citenamefont
  {Beenakker}}]{Choy2011}%
  \BibitemOpen
  \bibfield  {author} {\bibinfo {author} {\bibfnamefont {T.-P.}\ \bibnamefont
  {Choy}}, \bibinfo {author} {\bibfnamefont {J.~M.}\ \bibnamefont {Edge}},
  \bibinfo {author} {\bibfnamefont {A.~R.}\ \bibnamefont {Akhmerov}}, \ and\
  \bibinfo {author} {\bibfnamefont {C.~W.~J.}\ \bibnamefont {Beenakker}},\
  }\href {\doibase 10.1103/PhysRevB.84.195442} {\bibfield  {journal} {\bibinfo
  {journal} {Phys. Rev. B}\ }\textbf {\bibinfo {volume} {84}},\ \bibinfo
  {pages} {195442} (\bibinfo {year} {2011})}\BibitemShut {NoStop}%
\bibitem [{\citenamefont {Nadj-Perge}\ \emph {et~al.}(2013)\citenamefont
  {Nadj-Perge}, \citenamefont {Drozdov}, \citenamefont {Bernevig},\ and\
  \citenamefont {Yazdani}}]{Nadj-Perge:2013cr}%
  \BibitemOpen
  \bibfield  {author} {\bibinfo {author} {\bibfnamefont {S.}~\bibnamefont
  {Nadj-Perge}}, \bibinfo {author} {\bibfnamefont {I.~K.}\ \bibnamefont
  {Drozdov}}, \bibinfo {author} {\bibfnamefont {B.~A.}\ \bibnamefont
  {Bernevig}}, \ and\ \bibinfo {author} {\bibfnamefont {A.}~\bibnamefont
  {Yazdani}},\ }\href {\doibase 10.1103/PhysRevB.88.020407} {\bibfield
  {journal} {\bibinfo  {journal} {Phys. Rev. B}\ }\textbf {\bibinfo {volume}
  {88}},\ \bibinfo {pages} {020407} (\bibinfo {year} {2013})}\BibitemShut
  {NoStop}%
\bibitem [{\citenamefont {Kjaergaard}\ \emph {et~al.}(2012)\citenamefont
  {Kjaergaard}, \citenamefont {W\"{o}lms},\ and\ \citenamefont
  {Flensberg}}]{Kjaergaard:2012tg}%
  \BibitemOpen
  \bibfield  {author} {\bibinfo {author} {\bibfnamefont {M.}~\bibnamefont
  {Kjaergaard}}, \bibinfo {author} {\bibfnamefont {K.}~\bibnamefont
  {W\"{o}lms}}, \ and\ \bibinfo {author} {\bibfnamefont {K.}~\bibnamefont
  {Flensberg}},\ }\href {\doibase 10.1103/PhysRevB.85.020503} {\bibfield
  {journal} {\bibinfo  {journal} {Phys. Rev. B}\ }\textbf {\bibinfo {volume}
  {85}},\ \bibinfo {pages} {020503} (\bibinfo {year} {2012})}\BibitemShut
  {NoStop}%
\bibitem [{\citenamefont {Nadj-Perge}\ \emph {et~al.}(2014)\citenamefont
  {Nadj-Perge}, \citenamefont {Drozdov}, \citenamefont {Li}, \citenamefont
  {Chen}, \citenamefont {Jeon}, \citenamefont {Seo}, \citenamefont {MacDonald},
  \citenamefont {Bernevig},\ and\ \citenamefont {Yazdani}}]{Nadj-Perge:2014qv}%
  \BibitemOpen
  \bibfield  {author} {\bibinfo {author} {\bibfnamefont {S.}~\bibnamefont
  {Nadj-Perge}}, \bibinfo {author} {\bibfnamefont {I.~K.}\ \bibnamefont
  {Drozdov}}, \bibinfo {author} {\bibfnamefont {J.}~\bibnamefont {Li}},
  \bibinfo {author} {\bibfnamefont {H.}~\bibnamefont {Chen}}, \bibinfo {author}
  {\bibfnamefont {S.}~\bibnamefont {Jeon}}, \bibinfo {author} {\bibfnamefont
  {J.}~\bibnamefont {Seo}}, \bibinfo {author} {\bibfnamefont {A.~H.}\
  \bibnamefont {MacDonald}}, \bibinfo {author} {\bibfnamefont {B.~A.}\
  \bibnamefont {Bernevig}}, \ and\ \bibinfo {author} {\bibfnamefont
  {A.}~\bibnamefont {Yazdani}},\ }\href@noop {} {\bibfield  {journal} {\bibinfo
   {journal} {Science}\ }\textbf {\bibinfo {volume} {346}},\ \bibinfo {pages}
  {602} (\bibinfo {year} {2014})}\BibitemShut {NoStop}%
\bibitem [{\citenamefont {Ruby}\ \emph {et~al.}(2015)\citenamefont {Ruby},
  \citenamefont {Pientka}, \citenamefont {Peng}, \citenamefont {von Oppen},
  \citenamefont {Heinrich},\ and\ \citenamefont {Franke}}]{Ruby2015}%
  \BibitemOpen
  \bibfield  {author} {\bibinfo {author} {\bibfnamefont {M.}~\bibnamefont
  {Ruby}}, \bibinfo {author} {\bibfnamefont {F.}~\bibnamefont {Pientka}},
  \bibinfo {author} {\bibfnamefont {Y.}~\bibnamefont {Peng}}, \bibinfo {author}
  {\bibfnamefont {F.}~\bibnamefont {von Oppen}}, \bibinfo {author}
  {\bibfnamefont {B.~W.}\ \bibnamefont {Heinrich}}, \ and\ \bibinfo {author}
  {\bibfnamefont {K.~J.}\ \bibnamefont {Franke}},\ }\href {\doibase
  10.1103/PhysRevLett.115.197204} {\bibfield  {journal} {\bibinfo  {journal}
  {Phys. Rev. Lett.}\ }\textbf {\bibinfo {volume} {115}},\ \bibinfo {pages}
  {197204} (\bibinfo {year} {2015})}\BibitemShut {NoStop}%
\bibitem [{\citenamefont {{Pawlak}}\ \emph {et~al.}(2016)\citenamefont
  {{Pawlak}}, \citenamefont {{Kisiel}}, \citenamefont {{Klinovaja}},
  \citenamefont {{Meier}}, \citenamefont {{Kawai}}, \citenamefont {{Glatzel}},
  \citenamefont {{Loss}},\ and\ \citenamefont {{Meyer}}}]{Pawlak2016}%
  \BibitemOpen
  \bibfield  {author} {\bibinfo {author} {\bibfnamefont {R.}~\bibnamefont
  {{Pawlak}}}, \bibinfo {author} {\bibfnamefont {M.}~\bibnamefont {{Kisiel}}},
  \bibinfo {author} {\bibfnamefont {J.}~\bibnamefont {{Klinovaja}}}, \bibinfo
  {author} {\bibfnamefont {T.}~\bibnamefont {{Meier}}}, \bibinfo {author}
  {\bibfnamefont {S.}~\bibnamefont {{Kawai}}}, \bibinfo {author} {\bibfnamefont
  {T.}~\bibnamefont {{Glatzel}}}, \bibinfo {author} {\bibfnamefont
  {D.}~\bibnamefont {{Loss}}}, \ and\ \bibinfo {author} {\bibfnamefont
  {E.}~\bibnamefont {{Meyer}}},\ }\href {\doibase 10.1038/npjqi.2016.35}
  {\bibfield  {journal} {\bibinfo  {journal} {npj Quantum Information}\
  }\textbf {\bibinfo {volume} {2}},\ \bibinfo {eid} {16035} (\bibinfo {year}
  {2016})}\BibitemShut {NoStop}%
\bibitem [{\citenamefont {Ruby}\ \emph {et~al.}(2017)\citenamefont {Ruby},
  \citenamefont {Heinrich}, \citenamefont {Peng}, \citenamefont {von Oppen},\
  and\ \citenamefont {Franke}}]{Ruby2017}%
  \BibitemOpen
  \bibfield  {author} {\bibinfo {author} {\bibfnamefont {M.}~\bibnamefont
  {Ruby}}, \bibinfo {author} {\bibfnamefont {B.~W.}\ \bibnamefont {Heinrich}},
  \bibinfo {author} {\bibfnamefont {Y.}~\bibnamefont {Peng}}, \bibinfo {author}
  {\bibfnamefont {F.}~\bibnamefont {von Oppen}}, \ and\ \bibinfo {author}
  {\bibfnamefont {K.~J.}\ \bibnamefont {Franke}},\ }\href {\doibase
  10.1021/acs.nanolett.7b01728} {\bibfield  {journal} {\bibinfo  {journal}
  {Nano Letters}\ }\textbf {\bibinfo {volume} {17}},\ \bibinfo {pages} {4473}
  (\bibinfo {year} {2017})}\BibitemShut {NoStop}%
\bibitem [{\citenamefont {Klinovaja}\ \emph {et~al.}(2012)\citenamefont
  {Klinovaja}, \citenamefont {Stano},\ and\ \citenamefont
  {Loss}}]{Klinovaja:2012}%
  \BibitemOpen
  \bibfield  {author} {\bibinfo {author} {\bibfnamefont {J.}~\bibnamefont
  {Klinovaja}}, \bibinfo {author} {\bibfnamefont {P.}~\bibnamefont {Stano}}, \
  and\ \bibinfo {author} {\bibfnamefont {D.}~\bibnamefont {Loss}},\ }\href
  {\doibase 10.1103/PhysRevLett.109.236801} {\bibfield  {journal} {\bibinfo
  {journal} {Phys. Rev. Lett.}\ }\textbf {\bibinfo {volume} {109}},\ \bibinfo
  {pages} {236801} (\bibinfo {year} {2012})}\BibitemShut {NoStop}%
\bibitem [{\citenamefont {Pientka}\ \emph {et~al.}(2013)\citenamefont
  {Pientka}, \citenamefont {Glazman},\ and\ \citenamefont {von
  Oppen}}]{Pientka:2013}%
  \BibitemOpen
  \bibfield  {author} {\bibinfo {author} {\bibfnamefont {F.}~\bibnamefont
  {Pientka}}, \bibinfo {author} {\bibfnamefont {L.~I.}\ \bibnamefont
  {Glazman}}, \ and\ \bibinfo {author} {\bibfnamefont {F.}~\bibnamefont {von
  Oppen}},\ }\href {\doibase 10.1103/PhysRevB.88.155420} {\bibfield  {journal}
  {\bibinfo  {journal} {Phys. Rev. B}\ }\textbf {\bibinfo {volume} {88}},\
  \bibinfo {pages} {155420} (\bibinfo {year} {2013})}\BibitemShut {NoStop}%
\bibitem [{\citenamefont {Rainis}\ \emph {et~al.}(2014)\citenamefont {Rainis},
  \citenamefont {Saha}, \citenamefont {Klinovaja}, \citenamefont {Trifunovic},\
  and\ \citenamefont {Loss}}]{Rainis:2014le}%
  \BibitemOpen
  \bibfield  {author} {\bibinfo {author} {\bibfnamefont {D.}~\bibnamefont
  {Rainis}}, \bibinfo {author} {\bibfnamefont {A.}~\bibnamefont {Saha}},
  \bibinfo {author} {\bibfnamefont {J.}~\bibnamefont {Klinovaja}}, \bibinfo
  {author} {\bibfnamefont {L.}~\bibnamefont {Trifunovic}}, \ and\ \bibinfo
  {author} {\bibfnamefont {D.}~\bibnamefont {Loss}},\ }\href {\doibase
  10.1103/PhysRevLett.112.196803} {\bibfield  {journal} {\bibinfo  {journal}
  {Phys. Rev. Lett.}\ }\textbf {\bibinfo {volume} {112}},\ \bibinfo {pages}
  {196803} (\bibinfo {year} {2014})}\BibitemShut {NoStop}%
\bibitem [{\citenamefont {Marra}\ and\ \citenamefont
  {Cuoco}(2017)}]{Marra2017}%
  \BibitemOpen
  \bibfield  {author} {\bibinfo {author} {\bibfnamefont {P.}~\bibnamefont
  {Marra}}\ and\ \bibinfo {author} {\bibfnamefont {M.}~\bibnamefont {Cuoco}},\
  }\href {\doibase 10.1103/PhysRevB.95.140504} {\bibfield  {journal} {\bibinfo
  {journal} {Phys. Rev. B}\ }\textbf {\bibinfo {volume} {95}},\ \bibinfo
  {pages} {140504} (\bibinfo {year} {2017})}\BibitemShut {NoStop}%
\bibitem [{\citenamefont {DeGottardi}\ \emph
  {et~al.}(2013{\natexlab{a}})\citenamefont {DeGottardi}, \citenamefont {Sen},\
  and\ \citenamefont {Vishveshwara}}]{DeGottardi:2013a}%
  \BibitemOpen
  \bibfield  {author} {\bibinfo {author} {\bibfnamefont {W.}~\bibnamefont
  {DeGottardi}}, \bibinfo {author} {\bibfnamefont {D.}~\bibnamefont {Sen}}, \
  and\ \bibinfo {author} {\bibfnamefont {S.}~\bibnamefont {Vishveshwara}},\
  }\href {\doibase 10.1103/PhysRevLett.110.146404} {\bibfield  {journal}
  {\bibinfo  {journal} {Phys. Rev. Lett.}\ }\textbf {\bibinfo {volume} {110}},\
  \bibinfo {pages} {146404} (\bibinfo {year} {2013}{\natexlab{a}})}\BibitemShut
  {NoStop}%
\bibitem [{\citenamefont {DeGottardi}\ \emph
  {et~al.}(2013{\natexlab{b}})\citenamefont {DeGottardi}, \citenamefont
  {Thakurathi}, \citenamefont {Vishveshwara},\ and\ \citenamefont
  {Sen}}]{DeGottardi:2013b}%
  \BibitemOpen
  \bibfield  {author} {\bibinfo {author} {\bibfnamefont {W.}~\bibnamefont
  {DeGottardi}}, \bibinfo {author} {\bibfnamefont {M.}~\bibnamefont
  {Thakurathi}}, \bibinfo {author} {\bibfnamefont {S.}~\bibnamefont
  {Vishveshwara}}, \ and\ \bibinfo {author} {\bibfnamefont {D.}~\bibnamefont
  {Sen}},\ }\href {\doibase 10.1103/PhysRevB.88.165111} {\bibfield  {journal}
  {\bibinfo  {journal} {Phys. Rev. B}\ }\textbf {\bibinfo {volume} {88}},\
  \bibinfo {pages} {165111} (\bibinfo {year} {2013}{\natexlab{b}})}\BibitemShut
  {NoStop}%
\bibitem [{\citenamefont {Pérez}\ and\ \citenamefont
  {Martínez}(2017)}]{Perez:2017}%
  \BibitemOpen
  \bibfield  {author} {\bibinfo {author} {\bibfnamefont {M.}~\bibnamefont
  {Pérez}}\ and\ \bibinfo {author} {\bibfnamefont {G.}~\bibnamefont
  {Martínez}},\ }\href {http://stacks.iop.org/0953-8984/29/i=47/a=475503}
  {\bibfield  {journal} {\bibinfo  {journal} {Journal of Physics: Condensed
  Matter}\ }\textbf {\bibinfo {volume} {29}},\ \bibinfo {pages} {475503}
  (\bibinfo {year} {2017})}\BibitemShut {NoStop}%
\bibitem [{\citenamefont {Hoffman}\ \emph {et~al.}(2016)\citenamefont
  {Hoffman}, \citenamefont {Klinovaja},\ and\ \citenamefont
  {Loss}}]{Hoffman:2016}%
  \BibitemOpen
  \bibfield  {author} {\bibinfo {author} {\bibfnamefont {S.}~\bibnamefont
  {Hoffman}}, \bibinfo {author} {\bibfnamefont {J.}~\bibnamefont {Klinovaja}},
  \ and\ \bibinfo {author} {\bibfnamefont {D.}~\bibnamefont {Loss}},\ }\href
  {\doibase 10.1103/PhysRevB.93.165418} {\bibfield  {journal} {\bibinfo
  {journal} {Phys. Rev. B}\ }\textbf {\bibinfo {volume} {93}},\ \bibinfo
  {pages} {165418} (\bibinfo {year} {2016})}\BibitemShut {NoStop}%
\bibitem [{\citenamefont {Tezuka}\ and\ \citenamefont
  {Kawakami}(2012)}]{Tezuka:2012}%
  \BibitemOpen
  \bibfield  {author} {\bibinfo {author} {\bibfnamefont {M.}~\bibnamefont
  {Tezuka}}\ and\ \bibinfo {author} {\bibfnamefont {N.}~\bibnamefont
  {Kawakami}},\ }\href {\doibase 10.1103/PhysRevB.85.140508} {\bibfield
  {journal} {\bibinfo  {journal} {Phys. Rev. B}\ }\textbf {\bibinfo {volume}
  {85}},\ \bibinfo {pages} {140508} (\bibinfo {year} {2012})}\BibitemShut
  {NoStop}%
\bibitem [{\citenamefont {P\"oyh\"onen}\ \emph {et~al.}(2014)\citenamefont
  {P\"oyh\"onen}, \citenamefont {Weststr\"om}, \citenamefont {R\"ontynen},\
  and\ \citenamefont {Ojanen}}]{Poyhonen:2014}%
  \BibitemOpen
  \bibfield  {author} {\bibinfo {author} {\bibfnamefont {K.}~\bibnamefont
  {P\"oyh\"onen}}, \bibinfo {author} {\bibfnamefont {A.}~\bibnamefont
  {Weststr\"om}}, \bibinfo {author} {\bibfnamefont {J.}~\bibnamefont
  {R\"ontynen}}, \ and\ \bibinfo {author} {\bibfnamefont {T.}~\bibnamefont
  {Ojanen}},\ }\href {\doibase 10.1103/PhysRevB.89.115109} {\bibfield
  {journal} {\bibinfo  {journal} {Phys. Rev. B}\ }\textbf {\bibinfo {volume}
  {89}},\ \bibinfo {pages} {115109} (\bibinfo {year} {2014})}\BibitemShut
  {NoStop}%
\bibitem [{\citenamefont {Zhang}\ and\ \citenamefont
  {Nori}(2016)}]{Zhang:2016}%
  \BibitemOpen
  \bibfield  {author} {\bibinfo {author} {\bibfnamefont {P.}~\bibnamefont
  {Zhang}}\ and\ \bibinfo {author} {\bibfnamefont {F.}~\bibnamefont {Nori}},\
  }\href {http://stacks.iop.org/1367-2630/18/i=4/a=043033} {\bibfield
  {journal} {\bibinfo  {journal} {New Journal of Physics}\ }\textbf {\bibinfo
  {volume} {18}},\ \bibinfo {pages} {043033} (\bibinfo {year}
  {2016})}\BibitemShut {NoStop}%
\bibitem [{\citenamefont {Moore}\ \emph {et~al.}(2018)\citenamefont {Moore},
  \citenamefont {Stanescu},\ and\ \citenamefont {Tewari}}]{Moore:2016}%
  \BibitemOpen
  \bibfield  {author} {\bibinfo {author} {\bibfnamefont {C.}~\bibnamefont
  {Moore}}, \bibinfo {author} {\bibfnamefont {T.~D.}\ \bibnamefont {Stanescu}},
  \ and\ \bibinfo {author} {\bibfnamefont {S.}~\bibnamefont {Tewari}},\ }\href
  {\doibase 10.1103/PhysRevB.97.165302} {\bibfield  {journal} {\bibinfo
  {journal} {Phys. Rev. B}\ }\textbf {\bibinfo {volume} {97}},\ \bibinfo
  {pages} {165302} (\bibinfo {year} {2018})}\BibitemShut {NoStop}%
\bibitem [{\citenamefont {Liu}\ \emph {et~al.}(2018)\citenamefont {Liu},
  \citenamefont {Sau},\ and\ \citenamefont {Das~Sarma}}]{Liu:2018}%
  \BibitemOpen
  \bibfield  {author} {\bibinfo {author} {\bibfnamefont {C.-X.}\ \bibnamefont
  {Liu}}, \bibinfo {author} {\bibfnamefont {J.~D.}\ \bibnamefont {Sau}}, \ and\
  \bibinfo {author} {\bibfnamefont {S.}~\bibnamefont {Das~Sarma}},\ }\href
  {\doibase 10.1103/PhysRevB.97.214502} {\bibfield  {journal} {\bibinfo
  {journal} {Phys. Rev. B}\ }\textbf {\bibinfo {volume} {97}},\ \bibinfo
  {pages} {214502} (\bibinfo {year} {2018})}\BibitemShut {NoStop}%
\bibitem [{\citenamefont {Plugge}\ \emph {et~al.}(2017)\citenamefont {Plugge},
  \citenamefont {Rasmussen}, \citenamefont {Egger},\ and\ \citenamefont
  {Flensberg}}]{Plugge:2017dg}%
  \BibitemOpen
  \bibfield  {author} {\bibinfo {author} {\bibfnamefont {S.}~\bibnamefont
  {Plugge}}, \bibinfo {author} {\bibfnamefont {A.}~\bibnamefont {Rasmussen}},
  \bibinfo {author} {\bibfnamefont {R.}~\bibnamefont {Egger}}, \ and\ \bibinfo
  {author} {\bibfnamefont {K.}~\bibnamefont {Flensberg}},\ }\href
  {http://stacks.iop.org/1367-2630/19/i=1/a=012001} {\bibfield  {journal}
  {\bibinfo  {journal} {New Journal of Physics}\ }\textbf {\bibinfo {volume}
  {19}},\ \bibinfo {pages} {012001} (\bibinfo {year} {2017})}\BibitemShut
  {NoStop}%
\bibitem [{\citenamefont {Karzig}\ \emph {et~al.}(2017)\citenamefont {Karzig},
  \citenamefont {Knapp}, \citenamefont {Lutchyn}, \citenamefont {Bonderson},
  \citenamefont {Hastings}, \citenamefont {Nayak}, \citenamefont {Alicea},
  \citenamefont {Flensberg}, \citenamefont {Plugge}, \citenamefont {Oreg},
  \citenamefont {Marcus},\ and\ \citenamefont {Freedman}}]{Karzig2017}%
  \BibitemOpen
  \bibfield  {author} {\bibinfo {author} {\bibfnamefont {T.}~\bibnamefont
  {Karzig}}, \bibinfo {author} {\bibfnamefont {C.}~\bibnamefont {Knapp}},
  \bibinfo {author} {\bibfnamefont {R.~M.}\ \bibnamefont {Lutchyn}}, \bibinfo
  {author} {\bibfnamefont {P.}~\bibnamefont {Bonderson}}, \bibinfo {author}
  {\bibfnamefont {M.~B.}\ \bibnamefont {Hastings}}, \bibinfo {author}
  {\bibfnamefont {C.}~\bibnamefont {Nayak}}, \bibinfo {author} {\bibfnamefont
  {J.}~\bibnamefont {Alicea}}, \bibinfo {author} {\bibfnamefont
  {K.}~\bibnamefont {Flensberg}}, \bibinfo {author} {\bibfnamefont
  {S.}~\bibnamefont {Plugge}}, \bibinfo {author} {\bibfnamefont
  {Y.}~\bibnamefont {Oreg}}, \bibinfo {author} {\bibfnamefont {C.~M.}\
  \bibnamefont {Marcus}}, \ and\ \bibinfo {author} {\bibfnamefont {M.~H.}\
  \bibnamefont {Freedman}},\ }\href {\doibase 10.1103/PhysRevB.95.235305}
  {\bibfield  {journal} {\bibinfo  {journal} {Phys. Rev. B}\ }\textbf {\bibinfo
  {volume} {95}},\ \bibinfo {pages} {235305} (\bibinfo {year}
  {2017})}\BibitemShut {NoStop}%
\bibitem [{\citenamefont {Knapp}\ \emph {et~al.}(2018)\citenamefont {Knapp},
  \citenamefont {Karzig}, \citenamefont {Lutchyn},\ and\ \citenamefont
  {Nayak}}]{Knapp:2018}%
  \BibitemOpen
  \bibfield  {author} {\bibinfo {author} {\bibfnamefont {C.}~\bibnamefont
  {Knapp}}, \bibinfo {author} {\bibfnamefont {T.}~\bibnamefont {Karzig}},
  \bibinfo {author} {\bibfnamefont {R.~M.}\ \bibnamefont {Lutchyn}}, \ and\
  \bibinfo {author} {\bibfnamefont {C.}~\bibnamefont {Nayak}},\ }\href
  {\doibase 10.1103/PhysRevB.97.125404} {\bibfield  {journal} {\bibinfo
  {journal} {Phys. Rev. B}\ }\textbf {\bibinfo {volume} {97}},\ \bibinfo
  {pages} {125404} (\bibinfo {year} {2018})}\BibitemShut {NoStop}%
\bibitem [{\citenamefont {Khaneja}\ \emph {et~al.}(2005)\citenamefont
  {Khaneja}, \citenamefont {Reiss}, \citenamefont {Kehlet}, \citenamefont
  {Schulte-Herbr\"{u}ggen},\ and\ \citenamefont {Glaser}}]{Khaneja:2005tw}%
  \BibitemOpen
  \bibfield  {author} {\bibinfo {author} {\bibfnamefont {N.}~\bibnamefont
  {Khaneja}}, \bibinfo {author} {\bibfnamefont {T.}~\bibnamefont {Reiss}},
  \bibinfo {author} {\bibfnamefont {C.}~\bibnamefont {Kehlet}}, \bibinfo
  {author} {\bibfnamefont {T.}~\bibnamefont {Schulte-Herbr\"{u}ggen}}, \ and\
  \bibinfo {author} {\bibfnamefont {S.~J.}\ \bibnamefont {Glaser}},\ }\href
  {\doibase 10.1016/j.jmr.2004.11.004} {\bibfield  {journal} {\bibinfo
  {journal} {Journal of Magnetic Resonance}\ }\textbf {\bibinfo {volume}
  {172}},\ \bibinfo {pages} {296 } (\bibinfo {year} {2005})}\BibitemShut
  {NoStop}%
\bibitem [{\citenamefont {Glaser}\ \emph {et~al.}(2015)\citenamefont {Glaser},
  \citenamefont {Boscain}, \citenamefont {Calarco}, \citenamefont {Koch},
  \citenamefont {K{\"o}ckenberger}, \citenamefont {Kosloff}, \citenamefont
  {Kuprov}, \citenamefont {Luy}, \citenamefont {Schirmer}, \citenamefont
  {Schulte-Herbr{\"u}ggen}, \citenamefont {Sugny},\ and\ \citenamefont
  {Wilhelm}}]{Glaser:2015ug}%
  \BibitemOpen
  \bibfield  {author} {\bibinfo {author} {\bibfnamefont {S.~J.}\ \bibnamefont
  {Glaser}}, \bibinfo {author} {\bibfnamefont {U.}~\bibnamefont {Boscain}},
  \bibinfo {author} {\bibfnamefont {T.}~\bibnamefont {Calarco}}, \bibinfo
  {author} {\bibfnamefont {C.~P.}\ \bibnamefont {Koch}}, \bibinfo {author}
  {\bibfnamefont {W.}~\bibnamefont {K{\"o}ckenberger}}, \bibinfo {author}
  {\bibfnamefont {R.}~\bibnamefont {Kosloff}}, \bibinfo {author} {\bibfnamefont
  {I.}~\bibnamefont {Kuprov}}, \bibinfo {author} {\bibfnamefont
  {B.}~\bibnamefont {Luy}}, \bibinfo {author} {\bibfnamefont {S.}~\bibnamefont
  {Schirmer}}, \bibinfo {author} {\bibfnamefont {T.}~\bibnamefont
  {Schulte-Herbr{\"u}ggen}}, \bibinfo {author} {\bibfnamefont {D.}~\bibnamefont
  {Sugny}}, \ and\ \bibinfo {author} {\bibfnamefont {F.~K.}\ \bibnamefont
  {Wilhelm}},\ }\href {\doibase 10.1140/epjd/e2015-60464-1} {\bibfield
  {journal} {\bibinfo  {journal} {The European Physical Journal D}\ }\textbf
  {\bibinfo {volume} {69}},\ \bibinfo {pages} {279} (\bibinfo {year}
  {2015})}\BibitemShut {NoStop}%
\bibitem [{\citenamefont {Koch}(2016)}]{Koch:2016ly}%
  \BibitemOpen
  \bibfield  {author} {\bibinfo {author} {\bibfnamefont {C.~P.}\ \bibnamefont
  {Koch}},\ }\href {http://stacks.iop.org/0953-8984/28/i=21/a=213001}
  {\bibfield  {journal} {\bibinfo  {journal} {Journal of Physics: Condensed
  Matter}\ }\textbf {\bibinfo {volume} {28}},\ \bibinfo {pages} {213001}
  (\bibinfo {year} {2016})}\BibitemShut {NoStop}%
\bibitem [{\citenamefont {Brif}\ \emph {et~al.}(2010)\citenamefont {Brif},
  \citenamefont {Chakrabarti},\ and\ \citenamefont {Rabitz}}]{RabitzReview}%
  \BibitemOpen
  \bibfield  {author} {\bibinfo {author} {\bibfnamefont {C.}~\bibnamefont
  {Brif}}, \bibinfo {author} {\bibfnamefont {R.}~\bibnamefont {Chakrabarti}}, \
  and\ \bibinfo {author} {\bibfnamefont {H.}~\bibnamefont {Rabitz}},\ }\href
  {http://stacks.iop.org/1367-2630/12/i=7/a=075008} {\bibfield  {journal}
  {\bibinfo  {journal} {New Journal of Physics}\ }\textbf {\bibinfo {volume}
  {12}},\ \bibinfo {pages} {075008} (\bibinfo {year} {2010})}\BibitemShut
  {NoStop}%
\bibitem [{\citenamefont {Motzoi}\ \emph {et~al.}(2009)\citenamefont {Motzoi},
  \citenamefont {Gambetta}, \citenamefont {Rebentrost},\ and\ \citenamefont
  {Wilhelm}}]{Motzoi:2009ly}%
  \BibitemOpen
  \bibfield  {author} {\bibinfo {author} {\bibfnamefont {F.}~\bibnamefont
  {Motzoi}}, \bibinfo {author} {\bibfnamefont {J.~M.}\ \bibnamefont
  {Gambetta}}, \bibinfo {author} {\bibfnamefont {P.}~\bibnamefont
  {Rebentrost}}, \ and\ \bibinfo {author} {\bibfnamefont {F.~K.}\ \bibnamefont
  {Wilhelm}},\ }\href {\doibase 10.1103/PhysRevLett.103.110501} {\bibfield
  {journal} {\bibinfo  {journal} {Phys. Rev. Lett.}\ }\textbf {\bibinfo
  {volume} {103}},\ \bibinfo {pages} {110501} (\bibinfo {year}
  {2009})}\BibitemShut {NoStop}%
\bibitem [{\citenamefont {Rahmani}\ \emph {et~al.}(2017)\citenamefont
  {Rahmani}, \citenamefont {Seradjeh},\ and\ \citenamefont
  {Franz}}]{Rahmani:2017fs}%
  \BibitemOpen
  \bibfield  {author} {\bibinfo {author} {\bibfnamefont {A.}~\bibnamefont
  {Rahmani}}, \bibinfo {author} {\bibfnamefont {B.}~\bibnamefont {Seradjeh}}, \
  and\ \bibinfo {author} {\bibfnamefont {M.}~\bibnamefont {Franz}},\ }\href
  {\doibase 10.1103/PhysRevB.96.075158} {\bibfield  {journal} {\bibinfo
  {journal} {Phys. Rev. B}\ }\textbf {\bibinfo {volume} {96}},\ \bibinfo
  {pages} {075158} (\bibinfo {year} {2017})}\BibitemShut {NoStop}%
\bibitem [{\citenamefont {J\"ager}\ \emph {et~al.}(2014)\citenamefont
  {J\"ager}, \citenamefont {Reich}, \citenamefont {Goerz}, \citenamefont
  {Koch},\ and\ \citenamefont {Hohenester}}]{jager2014}%
  \BibitemOpen
  \bibfield  {author} {\bibinfo {author} {\bibfnamefont {G.}~\bibnamefont
  {J\"ager}}, \bibinfo {author} {\bibfnamefont {D.~M.}\ \bibnamefont {Reich}},
  \bibinfo {author} {\bibfnamefont {M.~H.}\ \bibnamefont {Goerz}}, \bibinfo
  {author} {\bibfnamefont {C.~P.}\ \bibnamefont {Koch}}, \ and\ \bibinfo
  {author} {\bibfnamefont {U.}~\bibnamefont {Hohenester}},\ }\href {\doibase
  10.1103/PhysRevA.90.033628} {\bibfield  {journal} {\bibinfo  {journal} {Phys.
  Rev. A}\ }\textbf {\bibinfo {volume} {90}},\ \bibinfo {pages} {033628}
  (\bibinfo {year} {2014})}\BibitemShut {NoStop}%
\bibitem [{\citenamefont {Thouless}\ and\ \citenamefont
  {Kirkpatrick}(1981)}]{Thouless:1981wq}%
  \BibitemOpen
  \bibfield  {author} {\bibinfo {author} {\bibfnamefont {D.~J.}\ \bibnamefont
  {Thouless}}\ and\ \bibinfo {author} {\bibfnamefont {S.}~\bibnamefont
  {Kirkpatrick}},\ }\href {http://stacks.iop.org/0022-3719/14/i=3/a=007}
  {\bibfield  {journal} {\bibinfo  {journal} {Journal of Physics C: Solid State
  Physics}\ }\textbf {\bibinfo {volume} {14}},\ \bibinfo {pages} {235}
  (\bibinfo {year} {1981})}\BibitemShut {NoStop}%
\bibitem [{\citenamefont {Das~Sarma}\ \emph {et~al.}(2016)\citenamefont
  {Das~Sarma}, \citenamefont {Nag},\ and\ \citenamefont
  {Sau}}]{DasSarma:2016ad}%
  \BibitemOpen
  \bibfield  {author} {\bibinfo {author} {\bibfnamefont {S.}~\bibnamefont
  {Das~Sarma}}, \bibinfo {author} {\bibfnamefont {A.}~\bibnamefont {Nag}}, \
  and\ \bibinfo {author} {\bibfnamefont {J.~D.}\ \bibnamefont {Sau}},\ }\href
  {\doibase 10.1103/PhysRevB.94.035143} {\bibfield  {journal} {\bibinfo
  {journal} {Phys. Rev. B}\ }\textbf {\bibinfo {volume} {94}},\ \bibinfo
  {pages} {035143} (\bibinfo {year} {2016})}\BibitemShut {NoStop}%
\bibitem [{\citenamefont {Chiu}\ \emph {et~al.}(2016)\citenamefont {Chiu},
  \citenamefont {Teo}, \citenamefont {Schnyder},\ and\ \citenamefont
  {Ryu}}]{chiu2016}%
  \BibitemOpen
  \bibfield  {author} {\bibinfo {author} {\bibfnamefont {C.-K.}\ \bibnamefont
  {Chiu}}, \bibinfo {author} {\bibfnamefont {J.~C.~Y.}\ \bibnamefont {Teo}},
  \bibinfo {author} {\bibfnamefont {A.~P.}\ \bibnamefont {Schnyder}}, \ and\
  \bibinfo {author} {\bibfnamefont {S.}~\bibnamefont {Ryu}},\ }\href {\doibase
  10.1103/RevModPhys.88.035005} {\bibfield  {journal} {\bibinfo  {journal}
  {Rev. Mod. Phys.}\ }\textbf {\bibinfo {volume} {88}},\ \bibinfo {pages}
  {035005} (\bibinfo {year} {2016})}\BibitemShut {NoStop}%
\bibitem [{\citenamefont {Akhmerov}\ \emph {et~al.}(2011)\citenamefont
  {Akhmerov}, \citenamefont {Dahlhaus}, \citenamefont {Hassler}, \citenamefont
  {Wimmer},\ and\ \citenamefont {Beenakker}}]{Akhmerov:2011pi}%
  \BibitemOpen
  \bibfield  {author} {\bibinfo {author} {\bibfnamefont {A.~R.}\ \bibnamefont
  {Akhmerov}}, \bibinfo {author} {\bibfnamefont {J.~P.}\ \bibnamefont
  {Dahlhaus}}, \bibinfo {author} {\bibfnamefont {F.}~\bibnamefont {Hassler}},
  \bibinfo {author} {\bibfnamefont {M.}~\bibnamefont {Wimmer}}, \ and\ \bibinfo
  {author} {\bibfnamefont {C.~W.~J.}\ \bibnamefont {Beenakker}},\ }\href
  {\doibase 10.1103/PhysRevLett.106.057001} {\bibfield  {journal} {\bibinfo
  {journal} {Phys. Rev. Lett.}\ }\textbf {\bibinfo {volume} {106}},\ \bibinfo
  {pages} {057001} (\bibinfo {year} {2011})}\BibitemShut {NoStop}%
\bibitem [{\citenamefont {Fulga}\ \emph {et~al.}(2011)\citenamefont {Fulga},
  \citenamefont {Hassler}, \citenamefont {Akhmerov},\ and\ \citenamefont
  {Beenakker}}]{Fulga:2011qv}%
  \BibitemOpen
  \bibfield  {author} {\bibinfo {author} {\bibfnamefont {I.~C.}\ \bibnamefont
  {Fulga}}, \bibinfo {author} {\bibfnamefont {F.}~\bibnamefont {Hassler}},
  \bibinfo {author} {\bibfnamefont {A.~R.}\ \bibnamefont {Akhmerov}}, \ and\
  \bibinfo {author} {\bibfnamefont {C.~W.~J.}\ \bibnamefont {Beenakker}},\
  }\href {\doibase 10.1103/PhysRevB.83.155429} {\bibfield  {journal} {\bibinfo
  {journal} {Phys. Rev. B}\ }\textbf {\bibinfo {volume} {83}},\ \bibinfo
  {pages} {155429} (\bibinfo {year} {2011})}\BibitemShut {NoStop}%
\bibitem [{\citenamefont {Fulga}\ \emph {et~al.}(2012)\citenamefont {Fulga},
  \citenamefont {Hassler},\ and\ \citenamefont {Akhmerov}}]{Fulga:2012kq}%
  \BibitemOpen
  \bibfield  {author} {\bibinfo {author} {\bibfnamefont {I.~C.}\ \bibnamefont
  {Fulga}}, \bibinfo {author} {\bibfnamefont {F.}~\bibnamefont {Hassler}}, \
  and\ \bibinfo {author} {\bibfnamefont {A.~R.}\ \bibnamefont {Akhmerov}},\
  }\href {\doibase 10.1103/PhysRevB.85.165409} {\bibfield  {journal} {\bibinfo
  {journal} {Phys. Rev. B}\ }\textbf {\bibinfo {volume} {85}},\ \bibinfo
  {pages} {165409} (\bibinfo {year} {2012})}\BibitemShut {NoStop}%
\bibitem [{\citenamefont {Fisher}\ and\ \citenamefont
  {Lee}(1981)}]{Fisher:1981jt}%
  \BibitemOpen
  \bibfield  {author} {\bibinfo {author} {\bibfnamefont {D.~S.}\ \bibnamefont
  {Fisher}}\ and\ \bibinfo {author} {\bibfnamefont {P.~A.}\ \bibnamefont
  {Lee}},\ }\href {\doibase 10.1103/PhysRevB.23.6851} {\bibfield  {journal}
  {\bibinfo  {journal} {Phys. Rev. B}\ }\textbf {\bibinfo {volume} {23}},\
  \bibinfo {pages} {6851} (\bibinfo {year} {1981})}\BibitemShut {NoStop}%
\bibitem [{\citenamefont {Stone}\ and\ \citenamefont
  {Szafer}(1988)}]{Stone:1988lj}%
  \BibitemOpen
  \bibfield  {author} {\bibinfo {author} {\bibfnamefont {A.~D.}\ \bibnamefont
  {Stone}}\ and\ \bibinfo {author} {\bibfnamefont {A.}~\bibnamefont {Szafer}},\
  }\href {\doibase 10.1147/rd.323.0384} {\bibfield  {journal} {\bibinfo
  {journal} {IBM Journal of Research and Development}\ }\textbf {\bibinfo
  {volume} {32}},\ \bibinfo {pages} {384} (\bibinfo {year} {1988})}\BibitemShut
  {NoStop}%
\bibitem [{\citenamefont {Baranger}\ and\ \citenamefont
  {Stone}(1989)}]{Baranger:1989hc}%
  \BibitemOpen
  \bibfield  {author} {\bibinfo {author} {\bibfnamefont {H.~U.}\ \bibnamefont
  {Baranger}}\ and\ \bibinfo {author} {\bibfnamefont {A.~D.}\ \bibnamefont
  {Stone}},\ }\href {\doibase 10.1103/PhysRevB.40.8169} {\bibfield  {journal}
  {\bibinfo  {journal} {Phys. Rev. B}\ }\textbf {\bibinfo {volume} {40}},\
  \bibinfo {pages} {8169} (\bibinfo {year} {1989})}\BibitemShut {NoStop}%
\bibitem [{\citenamefont {Wimmer}(2008)}]{Wimmer:2008ve}%
  \BibitemOpen
  \bibfield  {author} {\bibinfo {author} {\bibfnamefont {M.}~\bibnamefont
  {Wimmer}},\ }\emph {\bibinfo {title} {Quantum transport in nanostructures:
  From computational concepts to spintronics in graphene and magnetic tunnel
  junctions}},\ \href@noop {} {Ph.D. thesis},\ \bibinfo  {school}
  {Universit\"{a}t Regensburg} (\bibinfo {year} {2008})\BibitemShut {NoStop}%
\bibitem [{\citenamefont {Potter}\ and\ \citenamefont
  {Lee}(2011)}]{Potter:2011sr}%
  \BibitemOpen
  \bibfield  {author} {\bibinfo {author} {\bibfnamefont {A.~C.}\ \bibnamefont
  {Potter}}\ and\ \bibinfo {author} {\bibfnamefont {P.~A.}\ \bibnamefont
  {Lee}},\ }\href {\doibase 10.1103/PhysRevB.83.184520} {\bibfield  {journal}
  {\bibinfo  {journal} {Phys. Rev. B}\ }\textbf {\bibinfo {volume} {83}},\
  \bibinfo {pages} {184520} (\bibinfo {year} {2011})}\BibitemShut {NoStop}%
\bibitem [{\citenamefont {Akhmerov}(2010)}]{Akhmerov:2010kl}%
  \BibitemOpen
  \bibfield  {author} {\bibinfo {author} {\bibfnamefont {A.~R.}\ \bibnamefont
  {Akhmerov}},\ }\href {\doibase 10.1103/PhysRevB.82.020509} {\bibfield
  {journal} {\bibinfo  {journal} {Phys. Rev. B}\ }\textbf {\bibinfo {volume}
  {82}},\ \bibinfo {pages} {020509} (\bibinfo {year} {2010})}\BibitemShut
  {NoStop}%
\bibitem [{\citenamefont {Prada}\ \emph {et~al.}(2012)\citenamefont {Prada},
  \citenamefont {San-Jose},\ and\ \citenamefont {Aguado}}]{Prada:2012}%
  \BibitemOpen
  \bibfield  {author} {\bibinfo {author} {\bibfnamefont {E.}~\bibnamefont
  {Prada}}, \bibinfo {author} {\bibfnamefont {P.}~\bibnamefont {San-Jose}}, \
  and\ \bibinfo {author} {\bibfnamefont {R.}~\bibnamefont {Aguado}},\ }\href
  {\doibase 10.1103/PhysRevB.86.180503} {\bibfield  {journal} {\bibinfo
  {journal} {Phys. Rev. B}\ }\textbf {\bibinfo {volume} {86}},\ \bibinfo
  {pages} {180503} (\bibinfo {year} {2012})}\BibitemShut {NoStop}%
\bibitem [{\citenamefont {Kells}\ \emph {et~al.}(2012)\citenamefont {Kells},
  \citenamefont {Meidan},\ and\ \citenamefont {Brouwer}}]{Kells:2012}%
  \BibitemOpen
  \bibfield  {author} {\bibinfo {author} {\bibfnamefont {G.}~\bibnamefont
  {Kells}}, \bibinfo {author} {\bibfnamefont {D.}~\bibnamefont {Meidan}}, \
  and\ \bibinfo {author} {\bibfnamefont {P.~W.}\ \bibnamefont {Brouwer}},\
  }\href {\doibase 10.1103/PhysRevB.86.100503} {\bibfield  {journal} {\bibinfo
  {journal} {Phys. Rev. B}\ }\textbf {\bibinfo {volume} {86}},\ \bibinfo
  {pages} {100503} (\bibinfo {year} {2012})}\BibitemShut {NoStop}%
\bibitem [{\citenamefont {Bagrets}\ and\ \citenamefont
  {Altland}(2012)}]{Bagrets:2012}%
  \BibitemOpen
  \bibfield  {author} {\bibinfo {author} {\bibfnamefont {D.}~\bibnamefont
  {Bagrets}}\ and\ \bibinfo {author} {\bibfnamefont {A.}~\bibnamefont
  {Altland}},\ }\href {\doibase 10.1103/PhysRevLett.109.227005} {\bibfield
  {journal} {\bibinfo  {journal} {Phys. Rev. Lett.}\ }\textbf {\bibinfo
  {volume} {109}},\ \bibinfo {pages} {227005} (\bibinfo {year}
  {2012})}\BibitemShut {NoStop}%
\bibitem [{\citenamefont {Vuik}\ \emph {et~al.}(2016)\citenamefont {Vuik},
  \citenamefont {Eeltink}, \citenamefont {Akhmerov},\ and\ \citenamefont
  {Wimmer}}]{Vuik:2016wc}%
  \BibitemOpen
  \bibfield  {author} {\bibinfo {author} {\bibfnamefont {A.}~\bibnamefont
  {Vuik}}, \bibinfo {author} {\bibfnamefont {D.}~\bibnamefont {Eeltink}},
  \bibinfo {author} {\bibfnamefont {A.~R.}\ \bibnamefont {Akhmerov}}, \ and\
  \bibinfo {author} {\bibfnamefont {M.}~\bibnamefont {Wimmer}},\ }\href
  {http://stacks.iop.org/1367-2630/18/i=3/a=033013} {\bibfield  {journal}
  {\bibinfo  {journal} {New Journal of Physics}\ }\textbf {\bibinfo {volume}
  {18}},\ \bibinfo {pages} {033013} (\bibinfo {year} {2016})}\BibitemShut
  {NoStop}%
\bibitem [{\citenamefont {Woods}\ \emph {et~al.}(2018)\citenamefont {Woods},
  \citenamefont {Stanescu},\ and\ \citenamefont {Das~Sarma}}]{Woods2018}%
  \BibitemOpen
  \bibfield  {author} {\bibinfo {author} {\bibfnamefont {B.~D.}\ \bibnamefont
  {Woods}}, \bibinfo {author} {\bibfnamefont {T.~D.}\ \bibnamefont {Stanescu}},
  \ and\ \bibinfo {author} {\bibfnamefont {S.}~\bibnamefont {Das~Sarma}},\
  }\href {\doibase 10.1103/PhysRevB.98.035428} {\bibfield  {journal} {\bibinfo
  {journal} {Phys. Rev. B}\ }\textbf {\bibinfo {volume} {98}},\ \bibinfo
  {pages} {035428} (\bibinfo {year} {2018})}\BibitemShut {NoStop}%
\bibitem [{\citenamefont {Antipov}\ \emph {et~al.}(2018)\citenamefont
  {Antipov}, \citenamefont {Bargerbos}, \citenamefont {Winkler}, \citenamefont
  {Bauer}, \citenamefont {Rossi},\ and\ \citenamefont {Lutchyn}}]{Antipov2018}%
  \BibitemOpen
  \bibfield  {author} {\bibinfo {author} {\bibfnamefont {A.~E.}\ \bibnamefont
  {Antipov}}, \bibinfo {author} {\bibfnamefont {A.}~\bibnamefont {Bargerbos}},
  \bibinfo {author} {\bibfnamefont {G.~W.}\ \bibnamefont {Winkler}}, \bibinfo
  {author} {\bibfnamefont {B.}~\bibnamefont {Bauer}}, \bibinfo {author}
  {\bibfnamefont {E.}~\bibnamefont {Rossi}}, \ and\ \bibinfo {author}
  {\bibfnamefont {R.~M.}\ \bibnamefont {Lutchyn}},\ }\href {\doibase
  10.1103/PhysRevX.8.031041} {\bibfield  {journal} {\bibinfo  {journal} {Phys.
  Rev. X}\ }\textbf {\bibinfo {volume} {8}},\ \bibinfo {pages} {031041}
  (\bibinfo {year} {2018})}\BibitemShut {NoStop}%
\bibitem [{\citenamefont {Mikkelsen}\ \emph {et~al.}(2018)\citenamefont
  {Mikkelsen}, \citenamefont {Kotetes}, \citenamefont {Krogstrup},\ and\
  \citenamefont {Flensberg}}]{Mikkelsen2018}%
  \BibitemOpen
  \bibfield  {author} {\bibinfo {author} {\bibfnamefont {A.~E.~G.}\
  \bibnamefont {Mikkelsen}}, \bibinfo {author} {\bibfnamefont {P.}~\bibnamefont
  {Kotetes}}, \bibinfo {author} {\bibfnamefont {P.}~\bibnamefont {Krogstrup}},
  \ and\ \bibinfo {author} {\bibfnamefont {K.}~\bibnamefont {Flensberg}},\
  }\href {\doibase 10.1103/PhysRevX.8.031040} {\bibfield  {journal} {\bibinfo
  {journal} {Phys. Rev. X}\ }\textbf {\bibinfo {volume} {8}},\ \bibinfo {pages}
  {031040} (\bibinfo {year} {2018})}\BibitemShut {NoStop}%
\bibitem [{\citenamefont {Takei}\ \emph {et~al.}(2013)\citenamefont {Takei},
  \citenamefont {Fregoso}, \citenamefont {Hui}, \citenamefont {Lobos},\ and\
  \citenamefont {Das~Sarma}}]{Takei:2013}%
  \BibitemOpen
  \bibfield  {author} {\bibinfo {author} {\bibfnamefont {S.}~\bibnamefont
  {Takei}}, \bibinfo {author} {\bibfnamefont {B.~M.}\ \bibnamefont {Fregoso}},
  \bibinfo {author} {\bibfnamefont {H.-Y.}\ \bibnamefont {Hui}}, \bibinfo
  {author} {\bibfnamefont {A.~M.}\ \bibnamefont {Lobos}}, \ and\ \bibinfo
  {author} {\bibfnamefont {S.}~\bibnamefont {Das~Sarma}},\ }\href {\doibase
  10.1103/PhysRevLett.110.186803} {\bibfield  {journal} {\bibinfo  {journal}
  {Phys. Rev. Lett.}\ }\textbf {\bibinfo {volume} {110}},\ \bibinfo {pages}
  {186803} (\bibinfo {year} {2013})}\BibitemShut {NoStop}%
\bibitem [{\citenamefont {Fatin}\ \emph {et~al.}(2016)\citenamefont {Fatin},
  \citenamefont {Matos-Abiague}, \citenamefont {Scharf},\ and\ \citenamefont
  {\ifmmode \check{Z}\else \v{Z}\fi{}uti\ifmmode~\acute{c}\else
  \'{c}\fi{}}}]{Fatin:2016}%
  \BibitemOpen
  \bibfield  {author} {\bibinfo {author} {\bibfnamefont {G.~L.}\ \bibnamefont
  {Fatin}}, \bibinfo {author} {\bibfnamefont {A.}~\bibnamefont
  {Matos-Abiague}}, \bibinfo {author} {\bibfnamefont {B.}~\bibnamefont
  {Scharf}}, \ and\ \bibinfo {author} {\bibfnamefont {I.}~\bibnamefont
  {\ifmmode \check{Z}\else \v{Z}\fi{}uti\ifmmode~\acute{c}\else \'{c}\fi{}}},\
  }\href {\doibase 10.1103/PhysRevLett.117.077002} {\bibfield  {journal}
  {\bibinfo  {journal} {Phys. Rev. Lett.}\ }\textbf {\bibinfo {volume} {117}},\
  \bibinfo {pages} {077002} (\bibinfo {year} {2016})}\BibitemShut {NoStop}%
\bibitem [{\citenamefont {{Maurer}}\ \emph {et~al.}(2018)\citenamefont
  {{Maurer}}, \citenamefont {{Gamble}}, \citenamefont {{Tracy}}, \citenamefont
  {{Eley}},\ and\ \citenamefont {{Lu}}}]{Maurer:2018}%
  \BibitemOpen
  \bibfield  {author} {\bibinfo {author} {\bibfnamefont {L.~N.}\ \bibnamefont
  {{Maurer}}}, \bibinfo {author} {\bibfnamefont {J.~K.}\ \bibnamefont
  {{Gamble}}}, \bibinfo {author} {\bibfnamefont {L.}~\bibnamefont {{Tracy}}},
  \bibinfo {author} {\bibfnamefont {S.}~\bibnamefont {{Eley}}}, \ and\ \bibinfo
  {author} {\bibfnamefont {T.~M.}\ \bibnamefont {{Lu}}},\ }\href@noop {}
  {\bibfield  {journal} {\bibinfo  {journal} {ArXiv e-prints}\ } (\bibinfo
  {year} {2018})},\ \Eprint {http://arxiv.org/abs/1801.03676} {arXiv:1801.03676
  [cond-mat.mes-hall]} \BibitemShut {NoStop}%
\bibitem [{\citenamefont {Pilozzi}\ \emph {et~al.}(2018)\citenamefont
  {Pilozzi}, \citenamefont {Farrelly}, \citenamefont {Marcucci},\ and\
  \citenamefont {Conti}}]{Pilozzi2018}%
  \BibitemOpen
  \bibfield  {author} {\bibinfo {author} {\bibfnamefont {L.}~\bibnamefont
  {Pilozzi}}, \bibinfo {author} {\bibfnamefont {F.~A.}\ \bibnamefont
  {Farrelly}}, \bibinfo {author} {\bibfnamefont {G.}~\bibnamefont {Marcucci}},
  \ and\ \bibinfo {author} {\bibfnamefont {C.}~\bibnamefont {Conti}},\ }\href
  {https://doi.org/10.1038/s42005-018-0058-8} {\bibfield  {journal} {\bibinfo
  {journal} {Communications Physics}\ }\textbf {\bibinfo {volume} {1}},\
  \bibinfo {pages} {57} (\bibinfo {year} {2018})}\BibitemShut {NoStop}%
\bibitem [{Note1()}]{Note1}%
  \BibitemOpen
  \bibinfo {note} {S. Boutin, J. Camirand Lemyre and I. Garate, \protect \href
  {http://dx.doi.org/10.5281/zenodo.1486048}{"Majorana bound state engineering
  via efficient real-space parameter optimization,"} (2018), source code.
  [http://doi.org/10.5281/zenodo.1486048]}\BibitemShut {NoStop}%
\bibitem [{\citenamefont {Rowland}\ and\ \citenamefont
  {Jones}(2012)}]{Rowland:2012vy}%
  \BibitemOpen
  \bibfield  {author} {\bibinfo {author} {\bibfnamefont {B.}~\bibnamefont
  {Rowland}}\ and\ \bibinfo {author} {\bibfnamefont {J.~A.}\ \bibnamefont
  {Jones}},\ }\href {\doibase 10.1098/rsta.2011.0361} {\bibfield  {journal}
  {\bibinfo  {journal} {Philosophical Transactions of the Royal Society of
  London A: Mathematical, Physical and Engineering Sciences}\ }\textbf
  {\bibinfo {volume} {370}},\ \bibinfo {pages} {4636} (\bibinfo {year}
  {2012})}\BibitemShut {NoStop}%
\bibitem [{Note2()}]{Note2}%
  \BibitemOpen
  \bibinfo {note} {As $\Phi $ is in general a complex number, the actual
  performance index should be $|\Phi |^2$~\cite {Rowland:2012vy}. While
  important in practice, this distinction does not influence the description of
  the algorithm and the analysis of its computational complexity.}\BibitemShut
  {Stop}%
\bibitem [{\citenamefont {Lewenkopf}\ and\ \citenamefont
  {Mucciolo}(2013)}]{Lewenkopf:2013kk}%
  \BibitemOpen
  \bibfield  {author} {\bibinfo {author} {\bibfnamefont {C.~H.}\ \bibnamefont
  {Lewenkopf}}\ and\ \bibinfo {author} {\bibfnamefont {E.~R.}\ \bibnamefont
  {Mucciolo}},\ }\href {\doibase 10.1007/s10825-013-0458-7} {\bibfield
  {journal} {\bibinfo  {journal} {Journal of Computational Electronics}\
  }\textbf {\bibinfo {volume} {12}},\ \bibinfo {pages} {203} (\bibinfo {year}
  {2013})}\BibitemShut {NoStop}%
\bibitem [{\citenamefont {Groth}\ \emph {et~al.}(2014)\citenamefont {Groth},
  \citenamefont {Wimmer}, \citenamefont {Akhmerov},\ and\ \citenamefont
  {Waintal}}]{Groth:2014qc}%
  \BibitemOpen
  \bibfield  {author} {\bibinfo {author} {\bibfnamefont {C.~W.}\ \bibnamefont
  {Groth}}, \bibinfo {author} {\bibfnamefont {M.}~\bibnamefont {Wimmer}},
  \bibinfo {author} {\bibfnamefont {A.~R.}\ \bibnamefont {Akhmerov}}, \ and\
  \bibinfo {author} {\bibfnamefont {X.}~\bibnamefont {Waintal}},\ }\href
  {http://stacks.iop.org/1367-2630/16/i=6/a=063065} {\bibfield  {journal}
  {\bibinfo  {journal} {New Journal of Physics}\ }\textbf {\bibinfo {volume}
  {16}},\ \bibinfo {pages} {063065} (\bibinfo {year} {2014})}\BibitemShut
  {NoStop}%
\bibitem [{\citenamefont {Umerski}(1997)}]{Umerski:1997tg}%
  \BibitemOpen
  \bibfield  {author} {\bibinfo {author} {\bibfnamefont {A.}~\bibnamefont
  {Umerski}},\ }\href {\doibase 10.1103/PhysRevB.55.5266} {\bibfield  {journal}
  {\bibinfo  {journal} {Phys. Rev. B}\ }\textbf {\bibinfo {volume} {55}},\
  \bibinfo {pages} {5266} (\bibinfo {year} {1997})}\BibitemShut {NoStop}%
\bibitem [{Note3()}]{Note3}%
  \BibitemOpen
  \bibinfo {note} {Although we consider here a site-independent operator, our
  calculation naturally extends to site-dependent local operators with $A
  \rightarrow A_j$.}\BibitemShut {Stop}%
\bibitem [{\citenamefont {Boutin}\ \emph {et~al.}(2017)\citenamefont {Boutin},
  \citenamefont {Andersen}, \citenamefont {Venkatraman}, \citenamefont
  {Ferris},\ and\ \citenamefont {Blais}}]{boutin:2016nr}%
  \BibitemOpen
  \bibfield  {author} {\bibinfo {author} {\bibfnamefont {S.}~\bibnamefont
  {Boutin}}, \bibinfo {author} {\bibfnamefont {C.~K.}\ \bibnamefont
  {Andersen}}, \bibinfo {author} {\bibfnamefont {J.}~\bibnamefont
  {Venkatraman}}, \bibinfo {author} {\bibfnamefont {A.~J.}\ \bibnamefont
  {Ferris}}, \ and\ \bibinfo {author} {\bibfnamefont {A.}~\bibnamefont
  {Blais}},\ }\href {\doibase 10.1103/PhysRevA.96.042315} {\bibfield  {journal}
  {\bibinfo  {journal} {Phys. Rev. A}\ }\textbf {\bibinfo {volume} {96}},\
  \bibinfo {pages} {042315} (\bibinfo {year} {2017})}\BibitemShut {NoStop}%
\bibitem [{\citenamefont {Fletcher}(2000)}]{Fletcher:2000uq}%
  \BibitemOpen
  \bibfield  {author} {\bibinfo {author} {\bibfnamefont {R.}~\bibnamefont
  {Fletcher}},\ }\href@noop {} {\emph {\bibinfo {title} {Practical Methods of
  Optimization}}},\ A Wiley-Interscience Publication\ (\bibinfo  {publisher}
  {John Wiley \& Sons},\ \bibinfo {year} {2000})\BibitemShut {NoStop}%
\bibitem [{\citenamefont {Byrd}\ \emph {et~al.}(1995)\citenamefont {Byrd},
  \citenamefont {Lu}, \citenamefont {Nocedal},\ and\ \citenamefont
  {Zhu}}]{Byrd:1995fv}%
  \BibitemOpen
  \bibfield  {author} {\bibinfo {author} {\bibfnamefont {R.~H.}\ \bibnamefont
  {Byrd}}, \bibinfo {author} {\bibfnamefont {P.}~\bibnamefont {Lu}}, \bibinfo
  {author} {\bibfnamefont {J.}~\bibnamefont {Nocedal}}, \ and\ \bibinfo
  {author} {\bibfnamefont {C.}~\bibnamefont {Zhu}},\ }\href {\doibase
  10.1137/0916069} {\bibfield  {journal} {\bibinfo  {journal} {SIAM Journal on
  Scientific Computing}\ }\textbf {\bibinfo {volume} {16}},\ \bibinfo {pages}
  {1190} (\bibinfo {year} {1995})}\BibitemShut {NoStop}%
\bibitem [{Note4()}]{Note4}%
  \BibitemOpen
  \bibinfo {note} {As we consider a single subband nanowire, the number of
  local degree of freedoms is fixed to $M=4$ in all calculations. Hence,
  contrary to previous sections, we do not consider the value of $M$ in the
  complexity considerations. All methods considered should have a complexity
  $O(N^\xi M^3)$.}\BibitemShut {Stop}%
\bibitem [{Note5()}]{Note5}%
  \BibitemOpen
  \bibinfo {note} {The minigap is defined as the lowest excitation energy of
  the Majorana wire decoupled from the leads. In the main text, we refer to it
  as the \protect \emph {topological gap} or simply the \protect \emph {energy
  gap}.}\BibitemShut {Stop}%
\bibitem [{Note6()}]{Note6}%
  \BibitemOpen
  \bibinfo {note} {Matrix diagonalization is $O(N^3)$ and finite difference
  requires to perform $N$ of them for the problem considered.}\BibitemShut
  {Stop}%
\bibitem [{Note7()}]{Note7}%
  \BibitemOpen
  \bibinfo {note} {The exception is Fig.~\ref {mu_modulation}, where we take
  the parameters $\alpha = 0.05 t$, $\Delta =0.0225 t$ and $\protect \mathbf
  {B}_0/t = 0.027 \protect \mathaccentV {hat}05E\protect \mathbf
  {x}$.}\BibitemShut {Stop}%
\bibitem [{\citenamefont {Wan}\ \emph {et~al.}(2015)\citenamefont {Wan},
  \citenamefont {Kazakov}, \citenamefont {Manfra}, \citenamefont {Pfeiffer},
  \citenamefont {West},\ and\ \citenamefont {Rokhinson}}]{Wan:2015qv}%
  \BibitemOpen
  \bibfield  {author} {\bibinfo {author} {\bibfnamefont {Z.}~\bibnamefont
  {Wan}}, \bibinfo {author} {\bibfnamefont {A.}~\bibnamefont {Kazakov}},
  \bibinfo {author} {\bibfnamefont {M.~J.}\ \bibnamefont {Manfra}}, \bibinfo
  {author} {\bibfnamefont {L.~N.}\ \bibnamefont {Pfeiffer}}, \bibinfo {author}
  {\bibfnamefont {K.~W.}\ \bibnamefont {West}}, \ and\ \bibinfo {author}
  {\bibfnamefont {L.~P.}\ \bibnamefont {Rokhinson}},\ }\href@noop {} {\bibfield
   {journal} {\bibinfo  {journal} {Nature communications}\ }\textbf {\bibinfo
  {volume} {6}},\ \bibinfo {pages} {7426} (\bibinfo {year} {2015})}\BibitemShut
  {NoStop}%
\bibitem [{\citenamefont {Leung}\ \emph {et~al.}(2017)\citenamefont {Leung},
  \citenamefont {Abdelhafez}, \citenamefont {Koch},\ and\ \citenamefont
  {Schuster}}]{Leung:2017lh}%
  \BibitemOpen
  \bibfield  {author} {\bibinfo {author} {\bibfnamefont {N.}~\bibnamefont
  {Leung}}, \bibinfo {author} {\bibfnamefont {M.}~\bibnamefont {Abdelhafez}},
  \bibinfo {author} {\bibfnamefont {J.}~\bibnamefont {Koch}}, \ and\ \bibinfo
  {author} {\bibfnamefont {D.}~\bibnamefont {Schuster}},\ }\href {\doibase
  10.1103/PhysRevA.95.042318} {\bibfield  {journal} {\bibinfo  {journal} {Phys.
  Rev. A}\ }\textbf {\bibinfo {volume} {95}},\ \bibinfo {pages} {042318}
  (\bibinfo {year} {2017})}\BibitemShut {NoStop}%
\bibitem [{Note8()}]{Note8}%
  \BibitemOpen
  \bibinfo {note} {We consider again a scattering region of length $N=400$, and
  constrain the optimization problem to textures with a periodicity of $25$
  sites and $\protect \textbf {b}_j \cdot \protect \mathaccentV
  {hat}05E{\protect \mathbf {z}}=0$.}\BibitemShut {Stop}%
\end{thebibliography}
%

\end{document}